 \documentstyle[preprint,eqsecnum,aps,psfig,tighten]{revtex} % tighten format
\newcommand{\blm}[1]{{\bf #1}}
\newcommand{\sdoub}[2]{ {#1 \atopwithdelims() #2} }
\newcommand{\NPB}[3]{Nucl.~Phys.\ {\bf B#1}, #2 (19#3)}
\newcommand{\PLB}[3]{Phys.~Lett.\ {\bf B#1}, #2 (19#3)}
\newcommand{\PRD}[3]{Phys.~Rev.\ {\bf D#1}, #2 (19#3)}
\newcommand{\PRL}[3]{Phys.~Rev.~Lett.\ {\bf #1}, #2 (19#3)}
\newcommand{\ZPC}[3]{Z.~Phys.\ {\bf C#1}, #2 (19#3)}
\newcommand{\PTP}[3]{Prog.~Theor.~Phys.\ {\bf #1},  #2 (19#3)}
\newcommand{\MPLA}[3]{Mod.~Phys.~Lett.\ {\bf A#1}, #2 (19#3)}
\newcommand{\PR}[3]{Phys.~Rep.\ {\bf #1}, #2 (19#3)}

\newcommand{\EPJC}[3]{Eur.~Phys.~J.\ {\bf C#1}, #2 (19#3)}
\newenvironment{nstyl}{
\begin{list}{$\bullet$}
{\setlength{\itemsep}{0pt}
\setlength{\parsep}{0pt}}\relax}{\end{list}}

\begin{document}
%-----------------------------------------------------------------------------
%                            Definitions
%-----------------------------------------------------------------------------
\def\be{\begin{equation}}
\def\ee{\end{equation}}
\def\bea{\begin{eqnarray}}
\def\ena{\end{eqnarray}}
\def\belt{\begin{mathletters}}      % RevTex only
\def\enlt{\end{mathletters}}        % RevTex only
\def\symeq{=\kern-0.32em\raisebox{0.135ex}
          {$\stackrel{\mbox{.}}{\mbox{.}}\kern0.27em$}}
\def\cah{{\cal H}}
\def\bla{{\bf 1}}
\def\blb{{\bf 2}}
\def\blc{{\bf 3}}
\def\bnst{\begin{nstyl}}
\def\enst{\end{nstyl}}
\def\to{\rightarrow}
\def\dsty{\displaystyle}
%-----------------------------------------------------------------------------
%                             TITLE PAGE
%-----------------------------------------------------------------------------
\preprint { \vbox{ \hbox{hep-ph/9905278}  \hbox{IOA--05--99} } }
\title    { An efficient renormalization group  improved implementation  
            of the MSSM effective potential.  }
\author   { D. V. Gioutsos\thanks{ E-mail address: me00031@cc.uoi.gr }}
\address  { Division of Theoretical Physics, University of Ioannina \\
          Ioannina, GR - 451 10, GREECE }
\date     {\today}
\maketitle
%-----------------------------------------------------------------------------
\begin{abstract}
\indent
In the context of MSSM, a novel improving procedure based on the
renormalization group equation is applied to the effective potential in 
the Higgs sector. We focus on the one-loop radiative corrections computed 
in Landau gauge by using the mass independent renormalization scheme 
$\overline{\rm DR}$.  Thanks to the decoupling theorem, the well-known 
multimass scale problem is circumvented by switching to a new effective 
field theory every time a new particle threshold is encountered. We find 
that, for any field configuration, there is a convenient renormalization 
scale $\tilde{Q}^*$ at which the loop expansion respects the perturbation 
series hierarchy and the theory retains the vital property of stability.
\end{abstract}
%-----------------------------------------------------------------------------
%\medskip
%\pacs{PACS numbers: }
\newpage
%-----------------------------------------------------------------------------
%                           MAIN TEXT
%-----------------------------------------------------------------------------
\section{Introduction}
%-----------------------------------------------------------------------------
The standard model (SM) forms the bedrock of modern high-energy physics, 
and accurately describes all physical phenomena at energy scales up to 
$\sim 100$ GeV (electroweak scale). Even in the absence of a grand 
unification of strong and electroweak (EW) forces at a very high energy 
scale, it is clear that the SM must be modified to incorporate the effects 
of gravity at the Planck scale ($M_P \simeq 10^{19}$ GeV). In this context, 
the EW scale is put in by hand and there is no idea about its origin: it is 
a completely free input parameter. In a more complete theory one would 
like to understand its origin in terms of more fundamental 
parameters (e.g., $M_P$), but such a theory would need more structure than 
present in SM.

Moreover, in a general QFT containing an elementary scalar the mass of this 
scalar would be naturally at the scale of the cutoff of the theory\footnote
%-----------------------------------------------------------------------------
{If the SM were the full story, then the Higgs mass would be naturally of
${\cal O}(M_P)$.}
%-----------------------------------------------------------------------------
due to the quadratically divergent loop corrections to the scalar self 
energy (hierarchy problem \cite{gelde}). If one wants to protect the scalar 
masses from getting these large loop corrections, one needs to introduce a 
new symmetry.  The only known such symmetry is supersymmetry (SUSY) 
\cite{susy} which relates properties of bosons and fermions. 
Although there are various ways in which SUSY might be joined with the SM,
for simplicity one can pursue a {\em minimal\/} construction and attempt
to write down a Lagrangian, which is the most  general effective
Lagrangian invariant under SUSY transformations up to soft-breaking
terms. This then results in the Lagrangian of the Minimal Supersymmetric
Standard Model (MSSM) \cite{mssm}, where each SM particle is accompanied by 
a supersymmetric partner. Supersymmetry also requires two Higgs doublets,
as opposed to the single Higgs doublet of the SM. 

One of the most remarkable features of the MSSM is that it offers a
plausible senario for $SU(2)_L \times U(1)_Y$ symmetry breaking.
However, in this senario one has still to enforce 
phenomenologically a bounded from below potential and the absense of 
directions in field space that may induce a spontaneous breaking of
electric and/or color charge symmetries \cite{frer}
(a fact that clearly violates experimental observations).

From the theoretical point of view, in the SM electric and color charge
are certainly conserved in an automatical way, since the only fundamental
scalar field is the Higgs boson, a colorless electroweak doublet. The 
Higgs potential has a continuum of degenerate minima, but these are all
physically equivalent and one can always define the unbroken $U(1)_Y$
generator to be the electric charge. On the contrary,
in SUSY extensions of SM things become more complicated. Since the Higgs
sector contains at least two Higgs doublets $H_1, H_2$, one has to check
that the minimum of the Higgs potential still occurs for values of 
$H_1, H_2$ which are appropriately aligned in order to preserve the 
electric charge. Another perplexity arises from the fact that the 
supersymmetric theory (MSSM) has a large 
number of additional charged and color scalar fields, namely all the
sleptons $(\tilde{\ell})$ and squarks $(\tilde{q})$. Consequently,
conservation of color and electric charge symmetries requires
that the minimum of the whole potential 
$V(H_1, H_2, \tilde{\ell}, \tilde{q})$ still occurs at a point in the
field space where $\langle\tilde{q}\rangle, \langle \tilde{\ell} \rangle=0$
(realistic or true minimum).

Yet, the true effective potential in which the vacuum structure is 
encoded, is a poorly known object beyond the tree-level approximation.
One reason for this is the dependence of its loop corrections upon
the very many different mass scales present in MSSM, so that a 
renormalization group (RG) analysis becomes rather tricky. In general,
when one deals with a system possessing a large mass scale $Q_M$, compared 
with the scale $Q_{\mu}$ at which one discusses physics, large 
logarithms such as $\ln(Q_M/Q_{\mu})$ always appear which affect
the convergence of the perturbative realization of the potential (loop
expansion). In this situation, one considers resumming the perturbation
series by using the renormalization group equation.
Nonetheless, in many relistic applications one often has to deal with
an additional mass scale $Q_m$ with the hierarchy 
$Q_{\mu} \ll Q_m \ll Q_M$. In MSSM, for instance, one can regard
$Q_{\mu}, Q_m, Q_M$ as the weak, supersymmetry-breaking and unification
scales respectively. When we study such a system, we face the problem of
multimass scales \cite{multisc:1,multisc:2,multisc:3,multisc:4,msd}: 
There appear several types of logarithms 
$\ln(Q_M/Q_{\mu})$ and $\ln(Q_m/Q_{\mu})$, while we are able to sum up 
just a single logarithm by means of the RG equation.

But is it really necessary to take into account these obscure loop
corrections? Naively, one would argue that they cannot
change much of the qualitative pattern of the tree level minima.
On these grounds, let us recall that classical potential in MSSM receive 
contribution from 
three sources: D-terms, F-terms and soft-breaking terms. The first of
these provides the quartic terms $V_D = \lambda \varphi^4$ with 
$\lambda > 0$. Now along special directions in field space, known as 
D-flat directions, $V_D \sim 0$ can occur. If there is a minimum in
such a direction, then the change in the shape of the potential 
near the vicinity of that minimum, due to one-loop corrections, could 
create perturbatively a supplementary local minimum lower
than that already present at tree level.\footnote
%-----------------------------------------------------------------------------
{Provided that we keep under control the F-term contribution as well.} 
%-----------------------------------------------------------------------------
In other words, even if the one-loop corrected effective potential is 
expected to differ, point-to-point, only perturbatively from its tree level 
value, it is still possible that its shape be {\em locally\/} modified in 
such a way that new local minima (or at least stationary points) appear.

Clearly, a carefull RG improvement program is then essential in order
to deal with the physical problems already mentioned. Trying to 
illuminate this missing point,
in the present work a generalized improving procedure based on \cite{fws}
is applied to the MSSM effective potential. The main idea of the 
method is to make use of the decoupling theorem \cite{caraz}. By this 
theorem, it is made sufficient to treat essentially a single log factor at 
any renormalization scale, since all the heavy particles (heavier than that
scale) decouple and all the light particles (lighter than that scale)
yield more or less the same log factors.

The rest of the paper will be organized as follows. After setting our
notation and conventions, the state of the art concerning the EW
symmetry breaking in MSSM is briefly reviewed in Sec.~II\@. In the
next two sections, we describe the physical difficulties we have faced
in trying to extrapolate the well known low energy picture of MSSM to
higher energies and how they have been overcome. A detailed presentation 
of our numerical implementation is given in Secs.~V--VI, while 
Sec.~VII explains the reasoning behind our renormalization scale
choise. The final section is devoted to conclusions and further 
comments. Detailed formulae for the various field dependent MSSM mass 
eigenstates in a neutral Higgs background are presented in 
Appendix A and some special cases of them in Appendix B.
Finally, last two Appendices include technical details essential to
our numerical procedure, as well as a description of the numerical 
algorithms used for our purposes.
%-----------------------------------------------------------------------------
\section{Setting up the frame}
\subsection{The Lagrangian}
%-----------------------------------------------------------------------------
We are dealing with the MSSM i.e., the simplest supersymmetrized version of 
the SM. The requirements of minimal particle content and matter
parity conservation immediately dictate the expression of the 
$SU(3)_{C} \times SU(2)_{L} \times U(1)_{Y}$ invariant 
superpotential\footnote 
%-----------------------------------------------------------------------------
{$SU(2)$ indices are denoted by $i,j$, whereas $a$ is a color index and 
family indices are suppressed. Also $\epsilon_{12} = +1$.}
%-----------------------------------------------------------------------------
\be
\label{super}
{\cal W}=  \blm{Y_e} {\hat L}^j {\hat E}^c {\hat H}_1^i \epsilon_{ij}
         + \blm{Y_d} {\hat Q}^{ja} {\hat D}_a^c {\hat H}_1^i \epsilon_{ij}
         + \blm{Y_u} {\hat Q}^{ja} {\hat U}_a^c {\hat H}_2^i \epsilon_{ij}
         + \mu {\hat H}_1^i {\hat H}_2^j \epsilon_{ij}
\ee
where the chiral matter superfields transform as follows \\[3mm]
$\bullet$ Quark Superfields :
    ${\hat Q}=\sdoub{\hat{u}}{\hat{d}} \symeq (\blc,\blb,1/6)$, 
    ${\hat U}^c \symeq (\overline{\blc},\bla,-2/3)$,
    ${\hat D}^c \symeq (\overline{\blc},\bla,1/3)$  \\
$\bullet$ Lepton Superfields :
    ${\hat L}=\sdoub{\hat{\nu}}{\hat{\ell}} \symeq (\bla,\blb,-1/2)$, 
    ${\hat E}^c \symeq (\bla,\bla,1)$  \\
$\bullet$ Higgs Superfields :
    ${\hat H}_1=\sdoub{\hat{\cah}_1}{\hat{h}_1}\symeq(\bla,\blb,-1/2)$, 
    ${\hat H}_2=\sdoub{\hat{h}_2}{\hat{\cah}_2}\symeq(\bla,\blb,1/2)$ 
\\[3mm]
under the aforementioned gauge group.
Note that the free parameter $\mu$ and the $3 \times 3$ Yukawa matrices 
$\blm{Y_u}$, $\blm{Y_d}$, $\blm{Y_e}$ are generally complex.
These ingredients are enough to specify a globaly supersymmetric 
gauge invariant Lagrangian : ${\cal L}_{SUSY}$.

The fact that SUSY is not observed at low energies requires the 
introduction of extra supersymmetry breaking interactions. This breaking, 
however, should be such that no quadratic divergences appear and the 
technical ``solution'' to the hierarchy problem is not spoiled. Such 
terms are generally termed ``soft'' \cite{soft} and are interpreted as 
remnants of the spontaneous breaking of local SUSY in the underlying 
fundamental theory (Supergravity (SUGRA) \cite{sugra} or Superstrings
\cite{string}).
They include :
\bnst
\item Mass terms for all scalar fields.
\item Gaugino mass terms.
\item Biliner scalar interactions.
\item Trilinear scalar interactions.
\enst
Consequently, the most general SUSY breaking interaction Lagrangian 
with real mass terms, resulting from spontaneously broken SUGRA in 
the flat limit $(M_{P}{\rightarrow}{\infty})$ is\footnote 
%-----------------------------------------------------------------------------
{$\Gamma$, $R$ are $SU(2)$, $SU(3)$ group indices respectively 
and again we have supressed family indices.}
%-----------------------------------------------------------------------------
\bea
{\cal L}_{SOFT} 
        &=& -{1 \over 2} M_1 (\tilde{B} \tilde{B}) 
            -{1 \over 2} M_2 (\tilde{W^{\Gamma}} \tilde{W^{\Gamma}}) 
            -{1 \over 2} M_3 (\tilde{G^R} \tilde{G^R}) \nonumber \\
        & & -  m_{H_1}^2 |H_1|^2
            -m_{H_2}^2 |H_2|^2 -m^2_{\tilde{Q}}  |\tilde{Q}|^2
            -m^2_{\tilde{D}^c}  |\tilde{D}^c|^2
            - m^2_{\tilde{U}^c}  |\tilde{U}^c|^2  \nonumber\\
        & & - m^2_{\tilde{L}} |\tilde{L}|^2
            -m^2_{\tilde{E}^c}  |\tilde{E}^c|^2
            -(\blm{h_e}  \tilde{L}^j \tilde{E}^c H_1^i \epsilon_{ij}
            + \blm{h_d}  \tilde{Q}^{ja}   \tilde{D}_a^c H_1^i \epsilon_{ij}
            \nonumber\\
        & & + \blm{h_u}  \tilde{Q}^{ja} \tilde{U}_a^c H_2^i \epsilon_{ij} 
               + \mbox{H.c.})
            -(B \mu H_1^i H_2^j \epsilon_{ij}  + \mbox{H.c.})
\ena
We denote with $H_1$, $H_2$ the ordinary Higgs boson doublets and with
$\tilde{Q}$, $\tilde{D}^c$, $\tilde{U}^c$, $\tilde{L}$, $\tilde{E}^c$ 
the squark and slepton scalar fields. On the other hand, the 
gauginos $\tilde{B}$, $\tilde{W}$, $\tilde{G}$, are considered as 
two component Weyl spinors and $\blm{h} \equiv \blm{Y} \blm{A}$, 
where \blm{A} is a $3 \times 3$ matrix containing the ``soft'' trilinear 
scalar couplings. All extra soft parameters except masses are generally 
complex. This completes the picture for the low-energy effective theory. 

Altogether one would then need more than 100 real parameters to discribe
soft SUSY breaking in full generality. Clearly, some simplifying 
assumptions are necessary if we want to achieve something close to a 
complete study of parameter space. The following set of assumptions is 
adopted:
\bnst
\item We shall work in the approximation of vanishing intergenerational 
      mixing i.e., \\
      $ \blm{Y_e}=\mbox{diag}(Y_e^1,Y_e^2,Y_e^3), \quad
      \blm{Y_u}=\mbox{diag}(Y_u^1,Y_u^2,Y_u^3),  \quad
      \blm{Y_d}=\mbox{diag}(Y_d^1,Y_d^2,Y_d^3)    $ \\
      where all non-zero entries are real and positive.
\item $\mu$, $B$ and all trilinear soft couplings    
      are real.   \\
      (The phases of these parameters give large one-loop contributions
      to CP violating quantities, so practically they are quite 
      constrained \cite{phases}.) 
\item In our analysis we shall also keep Yukawas and
      trilinear soft couplings from light families,
      since their contributions to the
      one-loop effective potential (our main objective) are not 
      always negligible for an arbitrary field configuration.
\enst

A dramatic simplification of the structure of the SUSY-breaking 
interactions is provided either by grand unification assumptions
or by superstrings. The simplest possible choise arising from 
coupling the MSSM to minimal $N=1$ SUGRA is the following set of
assumptions:
\bnst
\item [(1)] Common gaugino mass : $m_{1/2}$  
\item [(2)] Common scalar mass : $m_0$  
\item [(3)] Common trilinear scalar coupling : $A_0$   
\enst
at a very large scale $M_X$.
More complicated alternatives also exist. However for the time 
being, for the sake of setting our notation, we shall focus on this
``universal'' senario.
 
This reduces the number of free parameters describing SUSY breaking 
to just four : The gaugino mass $m_{1/2}$, the scalar mass $m_0$, 
the trilinear and bilinear soft breaking parameters $A_0$ and $B$.
We do assume unification of the gauge couplings at scale 
$M_X \simeq 2 \times 10^{16}$ GeV.
At this point, no specific relation is assumed for the Yukawa couplings
at grand unification scale (GUT).\footnote{
%-----------------------------------------------------------------------------
From a theoretical perspective it would be more natural to impose 
boundary conditions at the (reduced) Planck scale 
$M_P / \sqrt{8\pi} \simeq 2.4\times10^{18}$ GeV
or perhaps the string compactification one $M_C \simeq
5 \times 10^{17}$ GeV \cite{dines}, rather than the GUT scale. 
In string theory, unification of the gauge couplings can be achieved 
even in the absence of a grand unification. In that case,
from our point of view, it is of no importance
whether we impose the above conditions at $M_X$ or $M_C$.}
%-----------------------------------------------------------------------------

In order to discuss the physical implications of this Lagrangian
at low-energy, we need to renormalize the relevant parameters 
at experimentally accesible energies
by computing them from a set of coupled renormalization group equations
(RGE) \cite{inoune}. (A detailed discussion of this procedure will be given 
in the following sections). This leads us to radiative gauge symmetry
breaking which we now consider.
%-----------------------------------------------------------------------------
\subsection{RG improvement and electroweak breaking}
%-----------------------------------------------------------------------------
The effective potential in general admits a loop expansion \cite{colm} as :
\be
   \hat{V}_{eff}(\lambda_{\alpha}, \phi; Q) = 
   \hat{V}_{eff}(\lambda_{\alpha}, 0;Q) +
   \sum_{n=1}^{\infty}{1 \over n!} \phi^n \Gamma^{(n)}(p_i=0)  
\ee
where $\hat{V}_{eff}(\lambda_{\alpha}, 0;Q)$ is a non-trivial function 
that receives contributions from all orders in perturbation theory,
$\lambda_{\alpha}$ are the couplings of the theory, $Q$ is a mass 
parameter upon which it explicitly depends and $\Gamma^{(n)}$ are the 
connected proper vertices at zero external momenta. 
Using a particular regularization scheme, one can get rid of the 
infinities by absorbing them into the renormalization of the basic 
parameters of the theory ($\lambda_{\alpha}, \phi$).
The resulting parameters become then $Q$ dependent and their evolution 
is determined by the beta functions $\beta_{\alpha}$ (couplings) and 
the anomalous dimensions $\gamma$ (fields). When a similar renormalization 
program is applied to $\hat{V}_{eff}$, order by order in perturbation
series, the resulting potential function ${\cal V}_{eff}$ can be cast
in a formally $Q$-dependent loopwise expansion
\be
\label{renv}
  {\cal V}_{eff}(\lambda_{\alpha}(Q),\phi(Q); Q) = \sum^{\infty}_{n=0}
     V^{(n)} = V^{(0)} + V^{(1)} + \ldots
\ee
Since $Q$ is an unphysical quantity, the potential in 
Eq.~(\ref{renv}) should not depend on 
the choise of it. This property can be achieved, if a change in the 
scale $Q$ can be compensated by appropriate changes in the couplings
and field rescalings. Mathematically it can be formulated by a
RG equation satisfied by the effective potential
\be
  \left( Q {d \over dQ} + 
   \beta_{\alpha} { \partial \over \partial \lambda_{\alpha}} -
   \gamma \phi { \partial \over \partial \phi} \right) V_{eff} = 0
\ee
which admits the solution 
\bea
  V_{eff} &=& V_{eff}(\lambda_{\alpha}(Q), \phi(Q);Q)   \nonumber \\
    &= & \Omega' + {\cal V}_{eff} \qquad\qquad\qquad 
                          (\Omega' : \mbox{ $\phi$ independent shift})
                           \nonumber \\
    &= & \Omega' + V^{(0)} + V^{(1)} + \ldots  
\ena
The point is that unless $\Omega'$ is specifically chosen, then $V_{eff}$
fails to satisfy the RG equation of the usual form 
\cite{multisc:3,omega:1,omega:2,omega:3}.
The obvious choise for $\Omega'$ is of course 
$\Omega' = - \sum_{n=0}^{\infty} V^{(n)}(\phi=0)$. From a practical
point of view, adding a field independent piece to $V_{eff}$ is 
perfectly harmless in problems where only field derivatives of 
$V_{eff}$ are of interest and a constant value is assigned to~$Q$.

An appealing feature of the MSSM is that it can lead to the breaking
of electroweak symmetry radiatively \cite{inoune,rad}. The correct 
$SU(2)_L \times U(1)_Y$ breaking down to $U(1)_{em}$ is achieved 
by restricting the vacuum expectation values (vevs) on the neutral 
Higgs manifold
\be
         \langle H^i_1 \rangle = \cah_1 \delta^i_1,
         ~~~~\langle H^i_2 \rangle = \cah_2 \delta^i_2 ,
     ~~~~\langle \tilde{q} \rangle=0,~~~~
         \langle \tilde{\ell} \rangle=0     
\label{vevs}
\ee
Here $\delta^i_j$, ($i,j = 1,2$) is the well known Kronecker symbol,
$\cah_1$, $\cah_2$ are taken real by gauge freedom and the last two 
equalities have to be satisfied by all the scalar quarks and leptons 
of the model. The low energy classical scalar potential along this 
direction is then
\be
     V^{(0)} = m_1^2 |\cah_1|^2 + m_2^2 |\cah_2|^2
   + 2 m_3^2 (\cah_1 \cah_2) + {g^2+g_2^2 \over 8}
    (|\cah_1|^2 - |\cah_2|^2)^2  
\label{tree}
\ee
where $m_1^2 = m_{H_1}^2 + \mu^2$, $m_2^2 = m_{H_2}^2 + \mu^2$, 
$m_3^2 = \mu B$ and $g$, $g_{2}$ are the $U(1)$ and $SU(2)$ gauge 
couplings.\footnote{ 
%-----------------------------------------------------------------------------
We are using the phase convention $\mu B < 0$, so a $\cah_1 \cah_2 > 0$
direction will ``deepen'' the potential.}
%-----------------------------------------------------------------------------

On the other hand, the one-loop effective Higgs potential of the model,
in Landau gauge using the $\overline{\rm DR}$ renormalization scheme 
\cite{siegel}, is
\bea
    V^{(1)} &=& {k \over 4}\;  Str\! \left\{
    {\cal M}^4 \left( 
    \ln{ {\cal M}^2 \over Q^2} - {3 \over 2}
    \right) \right\}   \nonumber \\
         &=& k \sum_{p \atop (M_p^2 \not= 0)} V^{(1)}_p
                    \label{oneloop} \\
     V^{(1)}_p &=& {(-1)^{2 S_p} \over 4} (2 S_p + 1) \, C_p \, {\cal N}_p
          \, M_p^4(\phi) \left( \ln{ |M_p^2(\phi)| \over Q^2}-{3 \over 2} 
     \right)    \nonumber
\ena
where $k=(16 \pi^2)^{-1}$ and ${\cal M}^2$ is the field dependent tree 
level squared mass matrix of the model. We denote by $M_p$ the mass 
eigenvalue of the ${\rm p}^{\rm th}$ particle and $S_p$, $C_p$ are its 
associated spin, color degrees of freedom.
${\cal N}_p$ is the number of its helicity states ($p$ runs over 
{\em all\/} particles), $\phi$ are the shifted scalar fields  and $Q$ 
is the renormalization scale. Finally, at one-loop order we have
\be
    \Omega' = - V^{(1)}(\lambda_{\alpha}(Q),\phi(Q) = 0;Q)  
\ee
so
\be
\label{used:pot}
   V_{1-loop} = \Omega' + V^{(0)} + V^{(1)} 
\ee
The parameters of the potential are taken as running ones, that is they 
vary with scale according to the two loop RGEs with one loop
thresholds in {\em all\/} running parameters \cite{sak1,2-loop:2}.

Before carrying on, we would like to comment on a subtle point in 
$ V^{(1)}$ definition. We know that at any acceptable minimum
all the eigenvalues of ${\cal M}^2$ must be positive, 
otherwise an imaginary part appears in $V^{(1)}$. This problem 
arises because of the non-convexity of the tree level potential 
$V^{(0)}$ \cite{multisc:2,convex}, when the gauge
symmetry is spontaneously broken. In our case, particularly in the
Higgs sector, if we compute the {\em tree\/} level field dependent 
mass eigenvalues for the values $(v_1^0,v_2^0)$ which minimize 
$V^{(0)}$ at low-energies, some of them would certainly vanish
(Goldstone modes). On the other hand, when we 
try to find the true minimum $(v_1,v_2)$ of $V_{1-loop}$, 
negative mass eigenvalues in the Higgs sector may now appear.
However, since these negative eigenvalues are of 
${\cal O}(\hbar)$, we can practically take the absolute value of 
$M_p^2$ inside the logarithm \cite{grz}.

Given the low energy scale of EW breaking we must use the 
renormalization group to evolve the parameters of the potential to a 
convenient scale such as $M_Z$ (physical Z-boson mass), where the 
experimental values of the gauge couplings are usually referred. In 
contrast to the tree level potential, $V_{1-loop}$ is relatively stable 
with respect to $Q$ around the electroweak scale 
\cite{omega:2,grz,stable,ramon}.
Therefore, the exact scale at which to minimize is no longer critical
as long as it is in the electroweak range. 

A glance at the superpotential (\ref{super}) shows that $H_2$ couples
to the $t$ (s)quarks, while $H_1$ only couples to $b$ (s)quarks and
$\tau$ (s)leptons by means of the Yukawa couplings. 
These Yukawa couplings enter the
RGEs for square scalar masses with positive sign. So their effect is 
always to decrease the Higgs masses as one evolves the RG equations 
downward from GUT scale to $M_Z$. Combined with the experimental fact 
that top quark is heavy, this can cause the RG-evolved $m_{H_2}^2$ to
run negative near the EW scale, helping to destabilize the point 
$\cah_1 = \cah_2 =0$ and so triggering the breaking of gauge symmetry.
If we define
\bea
   \overline{m}_i^2 &=& m_i^2 + \Sigma_i~~~~~(i=1,2)~~~~~~\mbox{where}
    \nonumber \\
    \Sigma_i   &=& {1 \over 2 \cah_i}\:
                    {\partial  \over \partial \cah_i}V^{(1)}
    \nonumber \\
        &=& {k \over 4}\sum_{p \atop (M_p^2 \not= 0)}
            (-1)^{2 S_p}  (2 S_p + 1) \, C_p \, {\cal N}_p
             {1 \over \cah_i}\: {\partial M_p^2 \over \partial \cah_i}\:
              f(M_p^2;Q)  
\ena
with
\bea
     f(x,Q) &=& x \left( \ln{|x| \over Q^2} -1 \right)
\ena
then the minimization of the potential yields the following two 
conditions\footnote
%-----------------------------------------------------------------------------
{For an analytic study of these conditions in the Higgs sector see
Ref.~\cite{moul}.}
%-----------------------------------------------------------------------------
among its parameters (all parameters are $Q$ {\em dependent\/})
\belt
\bea
\label{min1}
\sin 2\beta  &=&  -\frac{2 B \mu}{\overline{m}_1^2+\overline{m}_2^2}\\[3mm]
\label{min2}
\frac{1}{2} m_Z^2(M_Z)&=&\frac{\overline{m}_1^2-
\overline{m}_2^2 \tan^2\beta}{\tan^2\beta-1}
\ena
\enlt
where $m_Z^2(M_Z)$ is the running mass of the Z boson and 
$\tan\beta=v_2/v_1$. Note that in the above relations 
$\Sigma_i$ include contributions from {\em all\/} particles 
(see Appendix~A), $Q$ takes a {\em constant\/} value $(M_Z)$
and derivatives are taken with respect to
running fields $\cah_1(Q),\ \cah_2(Q)$, so there is no contribution 
from vacuum energy $(\partial \Omega' / \partial \cah_i(Q) = 0)$.
Moreover, since all the scalar fields ($\phi$) are multiplicative 
renormalizable, we conclude that if $\phi (Q_0)=0$, for some $Q_{0}$,
then $\phi (Q)=0$ for {\em every\/} $Q$. This is what actually happens 
for all scalar fields except Higgses (see Eq.~(\ref{vevs})).

Of course, we do not merely want 
$SU(2)_L \times U(1)_Y$ to be broken; we also want the Z boson
to have the experimentally measured mass ($M_Z$). Furthermore, in view 
of the strong dependence of some weak-scale quantities on $\tan\beta$,
it is often more convenient to treat $M_Z$ as an independent input
parameter. Since $m_Z^2 ={1 \over 2}(g^2+g_2^2) (v_1^2+v_2^2)$,
fixing $m_Z$ (equivalently $M_Z$) and $\tan\beta$ determines both
vevs $v_1,v_2$.

For all that, destabilizing the origin by a negative $({\rm mass})^2$ 
parameter is not enough to ensure the viability of the MSSM one loop 
scalar potential. We must also make sure that the potential
is bounded from below for arbitrarily large values of the scalar 
fields, so that Eq.~(\ref{used:pot}) will really have a minimum.
%-----------------------------------------------------------------------------
\section{Attempts to deal with high energies}
%-----------------------------------------------------------------------------
 Extending this well known low-energy picture to
high-energies (large field values), one is confronted with peculiar 
effective potential configurations. The simplest generalization is 
allowing Higgs fields, in Eqs.~(\ref{tree})--(\ref{oneloop}), to 
take arbitrary values keeping
at the same time the renormalization scale fixed at $M_Z$. However, this
assumption leads to an unbounded from below (UFB) potential. This 
realization is clearly physically undesirable. Before explaining the 
reason behind such a failure, let us see what causes this fake
instability.

Let $\cah_1=x_1$, $\cah_2=x_2$ and calculate all the field dependent 
mass eigenvalues in polar coordinates 
\be
     x_{1}=r\cos \theta, \qquad \qquad   x_{2}=r\sin \theta   
\ee
With no harm of generality, as shortly will be seen, we'll only take 
contributions from $3^{\rm rd}$ family. We intent to write $V_{1-loop}$
for $r \gg 1$. For our approximation to hold, coefficients that multiply
powers of $r$ should not be arbitrarily small. So, for each $r$, our 
approximation is valid only for those values of $\theta$ that respect
the above constraint. Choosing $\theta = \pi /2$ and using the 
formulae in Appendix B the potential becomes after some algebra
\belt
\be
    V_{1-loop}=V^{(0)}+ V^{(1)}+ \mbox{ (field independent piece)}   
\ee
where
\be
    V^{(0)}(r\gg 1) =  m_2^2 r^2 + \frac{g^2+g_2^2}{8} r^4 
     \simeq \frac{g^2+g_2^2}{8} r^4            
\ee
\be
   64 \pi^2 V^{(1)}(r\gg 1) = r^4 \left( 
       {\cal A}\, s(r) +
    \sum_{i=1}^{11} d_i U_i^2 \ln \frac{|U_i|}{Y_t^2} 
    \right)             
\ee
\be
   {\cal A} = \frac{2 g^4}{3} + \frac{(g^2-3g_2^2)^2}{24} - 
    3 Y_t^2(g^2+g_2^2) + \sum_{i=1}^9 d_i U_i^2   
\ee
\enlt
In the above, $s(r)=\ln( Y_t^2 r^2 / \tilde{Q}^2 )$, 
$\tilde{Q} = Q e^{3/4}$ and $d_i, U_i$ are given in Appendix B.
The crucial term for determining the behavior of the function presented
above is the logarithmic coefficient $({\cal A})$. This number due to 
large top 
Yukawa coupling is negative, so the whole function is UFB. (Note that
should we have taken contributions from light families, nothing would 
have changed since top quark Yukawa coupling still dominates).

Apparently, the main tool of our discussion is the loop expansion. So
ultimately one has to justify the convergence of loop expansion at 
high energies, ensuring in this way that only the first
terms in the series should be kept. Eventually, one has to study the 
logarithmic structure of the effective potential. It has been shown in
\cite{multisc:3,omega:1} that the L-loop effective potential contains 
logarithmic factors, only up to L-th power, whose magnitude control 
the convergence of the loop expansion. These factors have the general 
form 
\be
      s = \lambda \ln\frac{M^2(\phi_c)}{Q^2}  
\ee
where $\lambda$ is some coupling of the theory, $M$ is a field 
dependent mass eigenvalue in the presence of the background fileds 
$\phi_c$ and $Q$ is the renormalization scale.  In our case (MSSM), 
large field values generate large (field dependent) mass eigenvalues 
and obviously large log-factors ($Q$ is fixed at $M_Z$). So higher 
order corrections (2-loops etc.)\ become significant and should also be
taken into account. These corrections should raise the potential because 
the theory should be stable and can not have an UFB potential.
In conclusion, the one-loop approximation to the effective potential
renormalized at $M_Z$ for large field values 
{\itshape\bfseries is not reliable} and a different scale choise to 
control large logs is needed.

An alternative way of thinking about RG improvement is to view it as 
a reorganization of the perturbation series, in which the first term 
is the sum of all the {\em leading logarithms\/}, the second term 
represents the {\em next-to-leading logarithms}, and so on 
\cite{multisc:3,multisc:4,omega:1,omega:3}. 
The leading logarithms are terms of the form
$u^{(n)}\ln(M_1(\phi)/Q) \ln(M_2(\phi)/Q)\cdots 
\ln(M_n(\phi)/Q)$ and represent the most ``dangerous'' logarithmic 
terms at each order in perturbation theory.\footnote
%-----------------------------------------------------------------------------
{To compute the coefficient $u^{(n)}$ we need graphs with n-loops.
Furthermore, the next-to-leading logarithms are proportional to 
$u^{(n+1)}\ln(M_1(\phi)/Q) \ln(M_2(\phi)/Q)\cdots \ln(M_n(\phi)/Q)$ and
the tree level potential is counted as a leading logarithm.}
%-----------------------------------------------------------------------------
If we could fully sum the multiscale leading and
next-to-leading logs, we should have an approximation that is useful
dispite the existence of widely differing scales. However, when there
is more than one mass scales present it is less clear how to proceed;
no choise of $Q$ will destroy all the logarithms simoultaneously.

This point of view is adopted in the improvement prescription 
presented in \cite{multisc:3}. Let us briefly describe it.
Imagine a theory (like MSSM) having many coupling constants and mass
scales.  The problem in such multiscale cases is which log-factor
we should choose to sum up the leading, next-to-leading etc.\ logs. 
The best choise would be to take a particle whose coupling constant 
is the largest.\footnote{
%-----------------------------------------------------------------------------
In MSSM this role is played by the top quark.}
%-----------------------------------------------------------------------------
Calling that particle by label $k=0$ we have
\belt
\be
     s_0 = \lambda_0 \ln\frac{M_0^2(\phi_c)}{Q^2}   
\ee
\be
     s_k = \lambda_k \ln\frac{M_k^2(\phi_c)}{Q^2}=
    {\lambda_k \over \lambda_0}\,s_0 + u_k  \qquad \mbox{where} \qquad
   u_k = \lambda_k \ln\frac{M_k^2(\phi_c)} {M_0^2(\phi_c)}   
\ee
\enlt
If it happens all the masses $M_k^2(\phi_c)$ be of the same order 
independently of the background scalar fields 
$\phi_c$, then $u_k$ variables remain always of order 
${\cal O}(\lambda_k) \alt {\cal O}(\lambda_0)$. So all the log 
factors essentially can be treated as $s_0$ or 
$(\lambda_k/\lambda_0) s_0$,
since the differences $u_k$ are of order higher than the first $s_0$ 
term. In this case, the {\em field dependent\/} 
scale $ Q^{2}=M_0^2(\phi_c) \Leftrightarrow s_0=0$ correctly sums all
logs in the leading-log series expansion, improving thus  succesfully 
the effective potential.

We have numerically applied the prescription described above to MSSM 
and suprisingly found that with this scale prescription the effective
potential values near $x_1 =\pm x_2$ direction (H-diag) are quite
unexpected. H-diag direction, for large field values, 
develops saddle points (near 200 TeV! as shown in Fig.~1a) which lead 
to an unphysical (UFB) potential.  Unfortunately, near H-diag direction 
field dependent mass eigenvalues far away from origin are not all of the 
same order (see Appendix B). So some of the $u_k$ are large and one has 
to keep any higher powers of them in the leading-log series expansion. 
This in turn implies that higher loop corrections are significant and 
can not be neglected.  In other words, the absense of a unique scale 
choise (even field dependent), that eliminates large logs to all orders,
makes the improvement prescription of Ref.~\cite{multisc:3} for MSSM 
{\em unreliable}.

This deceptive deadlock of course steams from our careless treatment 
of different mass scales by a single scale parameter $Q$. Hiding all 
the heavy particle loop contributions in the redefinition of the low 
energy theory parameters, we can still solve the same RG equation for 
the effective potential by using different effective field theories.
This will be the subject of the next section.
%-----------------------------------------------------------------------------
\section{An alternative senario : V-thresholds}
%-----------------------------------------------------------------------------
Recently, the authors of Ref.~\cite{fws} have proposed a nice method to 
realize the attractive conjecture just mentioned.  Specifically, one 
should use the decoupling theorem to handle with the problem of many 
scales in the effective potential, using as decoupling scale for the 
mass eigenvalue $M^2(\phi)$ the scale 
\be
     \tilde{Q}^2 = |M^2(\phi)|
\ee
(recall that $\tilde{Q} = Q e^{3/4}$).
In other words, we use the following expansion as our impoved potenial
\belt
\be
     V^{(1)} = k \sum_i V_i^{(1)} \theta_i  
\ee
\be
   \theta_i \equiv \theta\Big(\tilde{Q}^2 - |M_i^2(\phi)|\Big)  
\ee
\be
    V_i^{(1)} = \frac{(-1)^{2 S_i}}{4} (2 S_i + 1)\, C_i\, {\cal N}_i\,
   M_i^4(\phi) \ln\frac{|M_i^2(\phi)|}{\tilde{Q}^2}  
\ee
\enlt
Despite the appearance of Heaviside functions the effective potential
is a {\em continuous\/} function (a discontinous potential is physically
meaningless). A discontinuity is likely to appear only at a decoupling
point (threshold). Let's examine what is happening when $\tilde{Q}^2$
approaches the $\ell^{\rm th}$ threshold. We have
\bnst
\item Just above the threshold :
      \[
      [V^{(1)}]_{\epsilon_+} = k \sum_{i=1}^{\ell -1}\,
      [V_i^{(1)} \theta_i]_{\epsilon_+} +
      k\, [V_{\ell}^{(1)} \theta_{\ell}]_{\epsilon_+}
      \]
\item Just below the threshold :
      \[
      [V^{(1)}]_{\epsilon_-}= k \sum_{i=1}^{\ell -1}\,
      [V_i^{(1)} \theta_i]_{\epsilon_-} 
      \]
\item At the threshold :
      \[
      [V^{(1)}]_{\epsilon=0} = k \sum_{i=1}^{\ell -1}\,
      [V_i^{(1)} \theta_i]_{\epsilon=0} + 
      k\, [V_{\ell}^{(1)} \theta_{\ell}]_{\epsilon=0}
      \]
\enst
But $V_{\ell}^{(1)} \propto M_{\ell}^4(\phi)
\ln( |M_{\ell}^2(\phi)| / \tilde{Q}^2 )$ and as 
$\tilde{Q}^2 \rightarrow |M_{\ell}^2(\phi)|$ 
(i.e., $\epsilon \rightarrow 0$), then 
$[V_{\ell}^{(1)} \theta_{\ell}]_{\epsilon=0} =0$ 
and $[V_{\ell}^{(1)} \theta_{\ell}]_{\epsilon_+} =0$
so, the potential function is continuous as it should be. 

On the other hand, defining
\be
\label{sigma-star}
   \Sigma_i^* = {1 \over 2 \cah_i} \sum_k 
   \frac{\partial V_k^{(1)}}{\partial \cah_i} \theta_k  
   \equiv \sum_k \Psi^{(k)}_i \theta_k
\ee
the corrections $\Sigma_i$ of Eq.~(2.12) become in this new approach
\be
   \Sigma_i = \Sigma_i^* - {1 \over 2 \cah_i} \sum_k V_k^{(1)}\,
    \delta\Big(\tilde{Q}^2 - |M_k^2(\phi)|\Big) \,
    {\partial |M_k^2| \over \partial \cah_i}       
\ee
The last term is obviously zero, so the stationary conditions have the well 
known form of Eqs.~(2.14a)--(2.14b) replacing $\Sigma_i$ with the new  
$\Sigma_i^*$ .

Let us recall that we are working in the one-loop approximation, which is
scale independent up to 2-loop corrections. These corrections, however,
may be very large due to the presence of potentially large logarithms
in $V^{(n>1)}$. Thus, one has to be careful in choosing a value of
Q which gives the finest approximation to $V_{1-loop}$. We'll come 
back to this critical issue during the numerical method's description,
but for completeness we stress here that in the above stated framework 
the potential is indeed bounded from below as Fig.~1b shows.
%-----------------------------------------------------------------------------
\section{Numerical procedure for radiative ew breaking}
%-----------------------------------------------------------------------------
In this section, following the analysis of \cite{sak1} we shall present
the procedure we use to restrict the SUSY parameter space to
those regions, where the requirement of radiative breaking of 
electroweak sector can be correctly implemented.

The generic problem, we have to tackle here, could be formulated as
a set of coupled {\em first-order\/} ordinary differential equations 
(ODEs) for the dependent variables (running parameters) $y_i$, 
$i=1,2,\ldots,n$ having the form 
\be
      y'_i(t) = f_i(y_1,y_2,\ldots,y_n)
\ee
where the functions $f_i$ on the right-hand side are known and 
$t \equiv 2 \ln(M_X/Q)$ is the independent variable (scale).
However, a problem involving ODEs is not completely specified by
its equations. Initial conditions (values of the running parameters
$y_i$) must also be supplied at some starting point $t_a$ which in
our case is: $t_a = 0$ i.e., $Q=M_X$. 
The underlying idea of any numerical routine, that solves an initial 
value problem, is to propagate the solution $y_i$ over an interval
$[t_a,t_b]$ with $t_b \not= t_a$, using a step-by-step calculation, 
which approximates values of the solution at discrete specified 
points in that interval.
As far as the MSSM case is concerned, we have used special routines from 
NAG fortran library (which is available to us) and particularly an 
Adams variable-order variale-step algorithm, which provides a choise 
of automatic error control and an option of a sophisticated 
root-finding technique \cite{adams}.

The initial conditions at $M_X$ scale will be choosen as simple as 
possible postponing for elsewhere the study of more complicated
alternatives. Thus, at the unification point taken to be 
${\cal O}(10^{16})$ GeV, we shall take 
\belt
\be
m_{\tilde{Q}}(M_X) =  m_{\tilde{D}^c}(M_X) = m_{\tilde{U}^c}(M_X) = m_0 
\ee
\be
      m_{\tilde{L}}(M_X) = m_{\tilde{E}^c}(M_X) = m_0  
\ee
\be
      m_{H_1}(M_X) = m_{H_2}(M_X) =  m_0        
\ee
and
\be
       M_1(M_X) = M_2(M_X) = M_3(M_X) = m_{1/2}       
\ee
In addition we take equal cubic couplings at $M_X$, i.e.,
\be
       A_e(M_X) = A_d(M_X) = A_u(M_X) = A_0     
\ee
\enlt
Obviously, all our ``universal'' boundary conditions are family blind.

Contact has to be made between the low-energy and high 
energy parameters of the theory \cite{low-susy}.  This will 
be achieved by integrating the RGEs from a superheavy scale, taken to 
be in the neighborhood of $10^{16}$ GeV, down to a scale $Q_0 \agt M_Z$.
In order to make a run starting from $M_X$ we must have the values of
all parameters at this scale, something difficult to achieve. The problem
is that the values of many parameters are known experimentally at low
scales. Also 1-loop stationary contitions are applied at such low 
energies. However, the values of other parameters, such as soft breaking
terms, are most easily understood at higher energies, where theoretical
simplification (e.g., universality) may be invoked. Thus, there is no 
scale at which both theoretical simplicity and experimental 
confirmation coexist.

Moreover, MSSM is characterized by two kinds of parameters, 
{\em constrained\/} and {\em free\/}. The former are constrained by 
experiment (gauge couplings, quark masses etc.) or by relations among 
themselves (stationary conditions at $M_Z$).
The latter, given the present experimental data, can not be constrained 
by these two critiria and must be viewed as input parameters.
These are $A_0$, $B_0$, $\mu_0$ (values at $M_X$), 
$m_{1/2}$, $m_0$, $\tan\beta$ (ratio of vevs at $M_Z$) and $M_t$ 
(physical mass for the top quark). 
In the present framework, $B_0$ and $\mu_0$ will be 
determined using the numerical integration routines in conjuction 
with the minimization of the one loop effective potential at $M_Z$ 
(i.e., Eqs. (2.14a)--(2.14b)).  Namely, stationary conditions\footnote
%-----------------------------------------------------------------------------
{We emphasize that $\Sigma_1^*$, $\Sigma_2^*$ involved in (2.14) 
contain contributions from {\em all\/} mass eigenvalues, even light ones.
These mass eigenvalues are obtained using everywhere
analytic formulae except for the neutralino sector where numerical 
diagonalization is performed (Appendix~A).}
%-----------------------------------------------------------------------------
at $M_Z$ will give $B(M_Z)$, $\mu(M_Z)$ and then we run back up to $M_X$ 
in order to determine $B_0$ and $\mu_0$. Note that the sign of $\mu$ is
not determined from (2.14a)--(2.14b), thus we must make a choise for it. 
So ultimately the free parameters of the theory are taken to be
\be
      A_0,~m_0,~m_{1/2},~\tan\beta,~M_t,~{\rm sign}(\mu)
\ee
On the other hand, the exact values of the constrained parameters at 
$M_X$ are affected by the choise of values for the free parameters 
at that scale, since the evolution of all parameters are coupled. 

We begin our numerical procedure at the EW scale which we take to be 
$M_Z$. This is an obvious choise, since many experimental quantities
are now available at that scale. 
At $Q=M_Z$ we take as input the well measured values of the Z mass
$M_Z= 91.1867 \pm 0.0020$ GeV, the electromagnetic coupling constant
$\alpha_{em}^{-1}(M_Z)|_{\overline{MS}} = 127.88 \pm 0.09$ \cite{pdata}
and the weak mixing angle 
$(\sin^2\theta_W)|_{\overline{MS}}=0.2316 - 0.88 \times 
10^{-7}(M_t^2-160^2){\rm GeV}^{-2}$ \cite{thesis}. 
In addition to the values of the gauge couplings\footnote
%-----------------------------------------------------------------------------
{The modified minimal subtraction ($\overline{\rm MS}$) values 
for the gauge couplings are related to their dimensional reduction
($\overline{\rm DR}$) ones through the
relations
$g_{\overline{MS}}=g_{\overline{DR}}\left(1 - \frac{Cg^2}{96\pi^2}\right)$
where $C=0,2,3$ respectively for the three gauge groups.} 
%-----------------------------------------------------------------------------
at $Q=M_Z$ one also 
needs the Yukawa couplings of quarks and leptons there. 
In order to determine $(Y_{\tau}, Y_b) \equiv (Y_e^3, Y_b^3)$
at $M_Z$, we run the gauge couplings $\alpha$ and $\alpha_3$
from their experimental values\footnote
%-----------------------------------------------------------------------------
{We start with the $\overline{\rm MS}$
values for the gauge couplings at $M_Z$ giving a trial input value for
the strong coupling constant $\alpha_3$ in the vicinity of 0.120.}
%-----------------------------------------------------------------------------
at $M_Z$ down to the $b$- and $\tau$-mass scales\footnote 
%-----------------------------------------------------------------------------
{For the $b$ quark and $\tau$ lepton our inputs are their physical 
masses: $M_b = 5.0$ GeV and \mbox{$M_{\tau}=1.777 $ GeV}.}
%-----------------------------------------------------------------------------
using three-loop QCD and two-loop QED RGEs \cite{SM-ramond}.
The relevant procedure is described with all details in \cite{sak1}.
Finally, the top Yukawa coupling ($Y_t$) is obtained from the running 
$t$-quark mass, which for our input value $M_t = 174$~GeV is 
$m_t(M_Z) \simeq  M_t / \left( 1+{5 \alpha_3 \over 3 \pi} \right) 
\simeq 162$ GeV.

The evolution of all couplings from $M_Z$ running upwards to higher 
energies determines the unification scale $M_X$ and the value of the 
unification coupling $\alpha_X$ (in $\overline{\rm DR}$) by 
\be
     \hat{\alpha}_1(M_X) = \hat{\alpha}_2(M_X)= \alpha_X 
\ee
Running down from $M_X$ to $M_Z$ the trial input value for $\alpha_3$
has now changed. The entire procedure is repeated several times. 
After one third of 
iterations is completed, we introduce finite part contributions to our 
low energy inputs such as QCD corrections to the top quark, gluino mass
and weak mixing angle $\sin^2 \theta_W |_{\overline{DR}}$. Also the gauge 
couplings values  $\hat{\alpha}_1$, $\hat{\alpha}_2$ 
(in $\overline{\rm DR}$) are now determined in terms of Fermi Constant 
$G_F = 1.16639 \times 10^{-5} \,{\rm GeV}^{-2}$, the Z boson mass and the 
electromagnetic coupling $\hat{\alpha}_{EM}^{-1} = 137.036$ \cite{pdata}. 
For details on these matters see Ref.~\cite{sak2}. The running $t$, $b$, 
$\tau$ quark masses at $M_Z$ are now defined through the
\be
   m_f(M_Z) = M_f + \Pi_f(Q=M_Z)  
\ee
where $M_f$ is the physical mass and $\Pi_f$ is the self energy 
corrections of the fermion.
This new procedure is now iterated until convergence is reached.
When convergence is established, the 
following actions are taken place at renormalization point $M_Z$ : 
\bnst
\item Light Yukawa couplings are being computed dividing the 
        fermion masses \cite{pdata} by the appropriate Higgs 
        vev. These couplings are considered as fixed with respect 
        to $Q$.\footnote
%-----------------------------------------------------------------------------
{Since $Y_{\ell ight} \simeq 0$ then  ${dY_{\ell ight} \over dt} =
(\cdots)Y_{\ell ight} \simeq 0$ thus 
$ Y_{\ell ight}(t) \simeq {\rm const.}$}
%-----------------------------------------------------------------------------
        On the other hand, the light trilinear scalar couplings 
        {\em do run}.
\item The whole set of the MSSM running parameters values 
        (gauge - Yukawa couplings, soft~masses - couplings and 
        Higgs vevs) is stored for later use (improvement of 
        effective potential around EW scale).
\item MSSM running parameters are evolved from their current values
        at $M_X$ to a lower scale $Q_{high}$\@. $Q_{high}$ is
        taken bigger than the largest field dependent mass 
        eigenvalue appearing in $V^{(1)}$, when the background 
        scalar fields are restricted in a hypercube of side 
        $S_{\infty}$. This point will be clarified later, but for the 
        moment we simply state that $Q_{high} \simeq 5 S_{\infty}$ 
        is a more or less satisfactory approximation for our 
        purposes.
\enst
Finally, our results should be reconciled with 
\bnst
\item Basic experimental constraints on supersymmetric masses 
      as well as Higgs bosons masses as shown in Table I \cite{pdata}.
\item Radiative electroweak breaking (existence of real non-zero vevs).
\item Positivity of eigenvalues of the mass matrix. Negative 
      eigenvalues mean that expanding around realistic minimum tachyonic
states appear making the true minimum of the theory unstable.
\enst
%-----------------------------------------------------------------------------
\section{Threshold parametrization for $\beta$-functions}
%-----------------------------------------------------------------------------
In the minimall low energy SUGRA model being considered, the 
super-particle spectrum is no longer degenerate. This should lead 
to various course corrections, each one occuring at the 
super-particle mass thresholds.
So the renormalization group $\beta$-functions must be cast in a
new form, which makes the implementation of the threshold effects
evident. Since the $\overline{\rm DR}$ RGEs are mass independent, each 
super-particle mass determines a boundary between two
effective theories. Above a particular mass threshold the associated
particle is present, whereas in the effective theory below the threshold
the particle is absent.

The simplest way to incorporate this is to treat the thresholds as steps
in the particle content of the renormalization group $\beta$-functions
\cite{sak1,ramon}. Let's briefly outline the procedure. Assume that $b$ 
is the beta function of a running parameter in the $\overline{\rm DR}$ 
scheme.  The corresponding RGE should be integrated from a superlarge 
scale $M_X$ down to any desirable value of $Q$. As we come down from 
$M_X$, as long as we are at scales larger than the heaviest particle in 
the spectrum, we include in $b$ contributions from all particles in the 
MSSM\@. When we cross the heaviest particle threshold, we switch in a new 
effective field theory with the heaviest particle integrated out and of 
course a new $b$. Coming further down in energy, we encounter the next 
particle threshold at which point we switch again to a new effective 
field theory with the two heaviest particles integrated out and a new $b$. 
That procedure goes on until all particles are exchausted.

Crossing a particle's threshold means that the renormalization scale 
has become smaller than its physical mass.
Hence, we need a condition to determine the exact point of 
decoupling (i.e., decoupling scale). For field configurations 
in the low energy regime ($\alt 300$ GeV) this is simply
\be
\label{soft}
      \tilde{Q}^2=m^2(Q),  \qquad    \tilde{Q} = Qe^{3/4} 
\ee
where $m^2(Q)$ is the running soft parameter corresponding to 
the particle.\footnote
%-----------------------------------------------------------------------------
{We use the factor $e^{3/4}$ for compatibility with the ``analogous'' 
decoupling in effective potential.}
%-----------------------------------------------------------------------------
Consequently, the step functions in RGEs will have the form 
\be
    \theta_m = \theta \Bigl(\tilde{Q}^2 - m^2(Q)\Bigr) 
\ee
Alternatively, for all other field configurations
we shall use a different condition to fix the decoupling point of a 
particle. 
Our condition now is
\be
   \tilde{Q}^2=|M^2(\phi;Q)|  
\ee
where $M^2$ is the field dependent mass eigenvalue for that 
particle. Analogously, the step functions in RGEs will become
\be
      \theta_M = \theta \Bigl(\tilde{Q}^2 - |M^2(\phi;Q)|\Bigr)   
\ee
 This procedure is generally more accurate than the approximation
stated in Eq.~(\ref{soft}), but in the case of true minimum it 
introduces non-trivial
field dependence, through mass eigenvalues, in $\Sigma_1^*$, $\Sigma_2^*$
and the simple stationary conditions (2.14) become extremely involved.
(For a presentation of the relevant $m^2$, $M^2$ see Appendix C.)
%-----------------------------------------------------------------------------
\section{Choosing the scale}
%-----------------------------------------------------------------------------
Our starting point is that the full effective potential is 
independent of the renormalization point $Q$ and thus satisfies 
the RGE
\be
   {d V_{eff} \over dQ} =0 
\ee
One can imediately write down the general solution to (7.1)
introducing the ``running'' distance $t$ from the initial values 
as : 
\belt
\be
    V_{eff} = V_{eff}(\lambda_{\alpha}(t),\phi(t);Q(t)) =
  \Omega' + V^{(0)} + V^{(1)} + {\cal O}(\hbar^2)  
\ee
\be
     Q(t) = M_X e^{-t/2},~~~~~~~~~~\Omega'= - V^{(1)}(\phi=0)  
\ee
\enlt
where $\lambda_a(t)$ are all dimensionless and dimensionfull couplings
of the MSSM and $\phi(t)=\zeta(t) \phi_c$ are the running fields with 
$\zeta(t)=\exp\{-\int_0^t dt'\gamma(t')\}$, $\gamma(t)$ being the 
anomalous dimension of the $\phi$ field.
The key to the usefulness of the RG is that we can choose a value of
$t$ such that the perturbation series converges more rapidly than the
series for $t=0$. Moreover, there is nothing to stop us choosing a
different value of $t$ for each value of $\phi$.

In order to validate the use of 1-loop effective potential one must
ensure that not only the couplings be perturbative, but that the 
loop expansion be convergent as well. In problems with only one mass 
scale RG improvement is straightforward. But for  the cases of 
interest here  there are several mass scales, so one must think up an 
improvement prescription. 
Moreover, the lack of analytic formulae describing the scale dependence 
of the quantities involved, as well as the absence of any profound
physical reasoning for choosing the appropriate scale make this task 
quite obscure. 

Earlier attempts \cite{omega:2,grz,stable,ramon,japan-improv} can not offer 
a substantial aid, since their object was an improvement in the low energy 
region.  For example, Ref.~\cite{grz} argues that there is a scale 
$\hat{Q}$ where one-loop stationary configuration coincides with the
tree level one. Thus by definition
\be
   \left. {\partial V^{(1)} \over \partial \cah_i} \right|_{\hat{Q}} = 0
\ee
The above definition is non-trivial to implement, so one should 
approximate
$\hat{Q}$ with an average of dominant field dependent masses.
For $|\cah_i| < 10^5$ GeV this is a legitimate approximation,
but when extending for $|\cah_i| \agt  10^5 $ GeV the potential 
develops an unphysical UFB escape along H-diag direction which 
forces us to search for something else.

Recently, another point of view have been introduced in Ref.~\cite{fws}.
Namely, one should compute at a scale $Q_R$ where both 
\belt
\be
\label{zerot}
     \left. V^{(1)} \right|_{Q_R}  \simeq 0 
\ee
\be
\label{fixsc}
   \left. {dV_{1-loop} \over dQ} \right|_{Q_R} 
           \simeq 0 
\ee 
\enlt
are satisfied. Clearly, for this $Q_R$ the 
radiative corrections to tree level are small and our approximation 
to the full potential has the least Q-dependence.
We have tried to implement the above prescription to MSSM, but the 
result is rather discouraging. The field dependent spectrum of MSSM
always consists of masses at the order of $M_Z$ (coming from the
neutralino sector). So application of (\ref{zerot}) will lead to a 
critical scale around $M_Z$, which as we saw gives a problematic 
potential.\footnote
%-----------------------------------------------------------------------------
{Of course this is not a real problem, since one can interpret
(\ref{zerot}) as an order of magnitude relation between zero and
first orders in loop expansion. So it is better to require
$|V^{(1)}| \ll |V^{(0)}|$ at $Q_R$. However, now there are several 
scale choises satisfying the previous condition.}
%-----------------------------------------------------------------------------
On the other hand, (\ref{fixsc}) seems much more reasonable
in the sense that the potential will be more scale independent.
However, in this case our numerical routines can not provide us with 
a scale for every field point $(\cah_1,\cah_2)$. Due to the step 
functions in RGEs the integration routine has to take too many 
intrinsic steps to preserve stability and accuracy of the solution. 
This in turn affects the performance {\em and\/} the numerical root 
finding facility when invoked for the equation (\ref{fixsc}).

To overcome this ambiguity in MSSM,
we shall make a physically motivated choise of scale which reproduces a
well behaved bounded from below effective potential one expects in a
stable theory.
Before carrying on, for clarity reasons, let us sketch briefly the
behaviour of $V^{(1)}$ near H-diag direction. Without harm of 
generality, we choose to examine the dominant contribution of 
top-stop sector. The corresponding partial sums of $V^{(1)}$ are 
\be
    \mbox{Top :} \quad P_t = -2\,{\cal C} M_t^4 
   \ln{|M_t^2| \over \tilde{Q}^2}\,\theta_t,~~~~~~\mbox{Stops :} \quad
   P_{T_i} = {\cal C} M_{T_i}^4\ln{|M_{T_i}^2| \over \tilde{Q}^2}
          \,\theta_{T_i}    
\ee
where ${\cal C}$ is some constant factor and $\theta_t,~\theta_{T_i}$
($i=1,2$) are the associated step functions. For non-zero partial sums 
we must take 
$\tilde{Q}> |M_t|,|M_{T_i}|$, i.e., $\ln{|M^2|\over \tilde{Q}^2}<0$
for all~$M$ involved ($P_t > 0$ and $P_{T_i} < 0$).
As we approach H-diag bottom points, the field dependent top-stop 
eigenvalues conspire to produce an extremely large\footnote 
%-----------------------------------------------------------------------------
{Compared to $V^{(0)}$.}
%-----------------------------------------------------------------------------
negative contribution to $V^{(1)}$ which is responsible for the 
UFB escape.

However, such a picture can not be reconciled with a perturbation series
hierarchy, neither with the notion of stability every acceptable 
physical theory should have.
Consequently, $\tilde{Q}$ should be taken such as top-stop and similar
``heavy'' pairs are decoupled, see Figs.~2a-b, making loop expansion 
meaningfull ($|V^{(0)}| \agt |V^{(1)}|$). One such rather conservative 
choise is $\tilde{Q^*} = 10^{-3} \sqrt{\cah_1^2+\cah_2^2}$ for large 
$\cah_i$ ($\cah_i \agt 10^8$ GeV).\footnote
%-----------------------------------------------------------------------------
{Dependence on $\cah_1,\cah_2$ is for practical reasons. 
$\tilde{Q^*}$ should be a smoothly continous function for the various 
field configurations and the only available ``free'' variables are 
these.}
%-----------------------------------------------------------------------------
Note that for this choise as $\cah_1,\cah_2$ approach $M_Z$, $Q^*$ 
will become lower than $M_Z$ making the RGE evolution ambiguous. 
The situation can be improved by defining 
\be
     \tilde{Q^*}= 10^{\omega(x)} \sqrt{\cah_1^2+\cah_2^2}
\ee
where $x={1 \over 2}\log{\cah_1^2+\cah_2^2 \over L^2}$ is the order 
of magnitude of a generalized ``radious'' in the field space and 
$L=1~{\rm GeV}$ makes the log argument dimensionless.

In the context of the approach we have followed, it seems rather non 
trivial to rigorously define $\omega(x)$. So one has to look for a 
qualitative fixing. Specifically, as $x$ decreases, field dependent 
masses decrease too, therefore $\tilde{Q}^*$ should decrease, 
otherwise the non-zero partial sums in $V^{(1)}$ could produce points 
deeper than the EW minimum. Using the cubic spline interpolation method 
\cite{stoer,num-rec} we can give an ansatz for $\omega(x)$, see 
Figs.~3a-b, compatible with continuity and the following constraints : 
(a) $Q^* \agt M_Z$, (b) $|V^{(0)}| \agt |V^{(1)}|$, 
(c) $\omega(9) \simeq -3$ in order to recover the previously stated 
conservative choise for $\tilde{Q}^*$, (d) as 
$\cah_i \to M_Z$, $Q^* \to M_Z$ i.e.,
\be
      \omega(x) = \cases{
         \omega_c - x                   &  if $\quad x \leq x_c$ \cr
 A_{11}(A_{12}+x)(A_{13}+x)(A_{14}+x)   &  if $\quad x_c < x \leq 4$ \cr
 A_{21}(A_{22}+x)(A_{23}+x)(A_{24}+x)   &  if $\quad 4   < x \leq 5$ \cr
 A_{31}(A_{32}+x)(A_{33}+A_{34} x+x^2)  &  if $\quad 5   < x \leq 6$ \cr
 A_{41}(A_{42}+x)(A_{43}+A_{44} x+x^2)  &  if $\quad 6   < x \leq 8$ \cr
 A_{51}(A_{52}+x)(A_{53}+A_{54} x+x^2)  &  if $\quad 8   < x       $ \cr  
       }
\ee
where by definition $x_c = \log(300\sqrt{2})$,
$\omega_c = \log({M_Z \over L} e^{3/4})$ and the coefficients $A$
are shown below
\be
A = \left(
\begin{array}{cccc}
    -0.691121  &  -4.17735  &  -4.       &  -2.39498  \\
     0.389965  &  -6.31201  &  -4.2182   &  -4.  \\
    -0.176436  &  -4.46928  &   37.1171  &  -11.5691  \\
     0.115471  &  -10.7737  &   27.1562  &  -10.2237  \\
    -0.0417777 &  -6.09931  &   183.38   &  -26.1999  \\
\end{array}
\right)
\ee

To complete this picture one also needs some ``boundary scale'' 
$Q_{high}$,
which shall provide the starting values of the running parameters 
for the evolution at $Q^*$. One convenient choise is an intermediate 
scale higher than the largest field dependent mass eigenvalue at the 
current field point. Specifically,
at some field point $\cah= \max(\cah_1,\cah_2)$ we approximately have 
$ \max_k\{\sqrt{|M_k^2(\cah_1,\cah_2)|}\} \simeq \cah $. Let 
$S_{\infty}$ denotes the upper bound order of magnitude of the allowed 
values for the fields. Then valid choises
for $Q_{high}$ are : $\tilde{Q}_{high} \agt S_{\infty}
\Rightarrow Q_{high} \agt S_{\infty} e^{3/4} \simeq 2.12 \, S_{\infty}$.
This is reasonable since the SM is supposed 
to be a low-energy effective theory coming from a more fundamental
theory. 

Specifying the scale $Q_{high}$ is not enough. We also need to know the
values of the running parameters and fields there. Since $Q_{high}$ is 
above all thresholds, the required values sould not depend on the background 
fields and the reasonable choise at an arbitrary field point is
\be
\label{ichigh}
    \lambda_{\alpha}(Q_{high};\cah_1(Q_{high}),\cah_2(Q_{high})) \equiv 
    \lambda_{\alpha}(Q_{high};v_1,v_2)   
\ee
where $v_1$, $v_2$ are the vevs of the EW minimum.\footnote 
%-----------------------------------------------------------------------------
{In general the value of a running parameter depend on the current
field point due to the field dependent thresholds involved in RGEs.}
%-----------------------------------------------------------------------------
In other words, to find the LHS of Eq.~(\ref{ichigh}) we simply
integrate the RGE from $M_X$ to this new $Q_{high}$ using as initial
conditions at $M_X$ the outcome of the iteration procedure described 
in Sec.~V\@. Evolving 
this set of values $\{\cah_1(Q_{high}), \cah_2(Q_{high}), 
\lambda_{\alpha}(Q_{high};v_1,v_2)\}$ to $Q^*$, using field dependent 
thresholds, the effective potential at the current field point can be 
constructed.
%-----------------------------------------------------------------------------
\section{Conclusions}
%-----------------------------------------------------------------------------
In the framework of MSSM, the fact that the top quark is heavy suggests an 
interesting possibility for explaining the spontaneous symmetry breaking at 
the EW scale, i.e., the radiative breaking senario. The key method to 
analyze such 
a senario is based on the RG equation. In describing the radiative 
symmetry breaking,
the most primitive approach is to use the tree level Higgs potential with
the RG running masses and couplings inserted.
There exists, however, a serious technical problem in finding symmetry 
breaking solutions. Namely, the results often depend badly on the choise
of the renormalization point $Q$ at which the RG running is terminated,
a fact that clearly reduces their reliability. 

As emphasized by the authors of 
Refs.\ \cite{omega:2,grz,stable,ramon,japan-improv}, 
addition of the one-loop corrections $V^{(1)}$ to the classical potential
ameliorates the scale dependence at least for low energies. However, we
have ascertained with surprise, that a careless treatment of these 
corrections for high energies formally jeopardizes the stability of the 
theory. Along a special direction (H-diag), where the magnitude of tree 
level potential $V^{(0)}$ is small, these corrections when carelessly 
treated predominate and produce an undesirable UFB escape. The reason behind 
such a failure must be sought in the inadequacy of a mass independent 
renormalization scheme $(\overline{\rm DR})$ to treat the very many mass 
scales of MSSM as one.\footnote
%-----------------------------------------------------------------------------
{For the case at hand (MSSM), the employnment of a mass dependent
renormalization scheme would be an arduous task (for an application to 
the simple Yukawa model see \cite{msd}).} 
%-----------------------------------------------------------------------------
The problem is that the decoupling of the various particles is not
automatically included in the formalism and has to be incorporated.

So to achieve our purpose we have tried to implement the decoupling
theorem in a manner proposed by the authors of Ref.~\cite{fws}.
A simple way to incorporate
it in the MSSM case is to treat the various particle thresholds as steps
in the $\beta$-functions, as well as in the one loop corrections of the
scalar potential. Each time we cross a threshold the $\beta$-function 
changes indicating that we have switched to a new effective field theory 
with the heaviest particle integrated out. At the same time, the associated 
particle's contribution to $V^{(1)}$ is dropped 
realizing in this way the process of decoupling. We stress here the
role played by the renormalization scale choice, as given in Sec.~VII\@. 
It should be wisely chosen in order to eliminate heavy particles whose 
participation induces a fake instability, while at the same time improve 
the convergence of the loop expansion (i.e., $|V^{(1)}| \alt |V^{(0)}|$).

Employing the framework just stated and using the ``merlin'' minimization 
program \cite{merlin} (for details see Appendix~D) we have scanned the 
dangerous H-diag direction in its entirety
in order to reveal unexpected local minima different than the true. This 
procedure has been repeated for a representative set of initial conditions
at $M_X$, but the outcome was always the same: a bounded from below potential
with a $SU(3)_C \times U(1)_Y$ symmetric minimum at EW scale.

Clearly, since the described method allows to undertake faraway excursions 
in the field space it can be utilized in less investigated situations.
For instance, in the past various authors have focused on conditions 
(involving soft trilinear scalar couplings), which are needed to ensure 
that a particular
SUSY model avoids an electric and/or color charge breaking ground state.
However, to our opinion a careful numerical analysis of the impact the 
one-loop corrections $V^{(1)}$ would have on these matters, is still 
required. Due to their inherent complexity and for the sake of presenting 
analytic expressions, one usually resorts to getting round contrivances 
in order to efficiently deal with the problem.
On the other hand, from our numerical point of view, we can directly
attack the problem of radiative corrections at the 
cost of loosing track of analytic elegance. These issues will be
the subject of a forthcoming publication \cite{g+v}.
%-----------------------------------------------------------------------------
%                          ACKNOWLEDGEMENTS 
%-----------------------------------------------------------------------------
\acknowledgments
%-----------------------------------------------------------------------------
It's a pleasure to thank Professor C.E.~Vayonakis for his advice, many 
suggestions, patient discussions, and a careful reading of the manuscript. 
We are greatful to K. Tamvakis for helpful discussions and important 
comments.  We also wish to thank A. Dedes for his collaboration during the 
earlier stages of this project and S.~Lola, I.~Lagaris, D.~Papageorgiou, 
J.~Rizos, D.E.~Katsanos for their invaluable help with the computer.
This work has been supported in part by program $\Pi{\rm ENE}\Delta~563$.
%-----------------------------------------------------------------------------
%                             APPENDIXES
%-----------------------------------------------------------------------------
%\newpage
\appendix
\section{Field dependent spectrum of MSSM}
%-----------------------------------------------------------------------------
\def\theequation{A.\arabic{equation}}
\setcounter{equation}{0}
%-----------------------------------------------------------------------------
We give here all the field dependent mass eigenstates of the MSSM in the 
presence of a neutral Higgs background and the necessary expressions 
$(\Psi_i)$ to compute $\Sigma_i^*$ of Eq.~(\ref{sigma-star}).
Similar formulae can also be found in Ref.~\cite{ar-nat}.
To simplify somehow the 
expressions we switch to the notation $\cah_1 = x_1,~\cah_2=x_2$.
The function $f(x)$, frequently used below, has already been defined in 
Sec.~II, as well as the rest of the notation.
\begin{description}
\item [$\bullet$ Gauge Bosons] $(W^{\pm},Z)$
      \be
      M^2(W^{\pm}) = {g_2^2 \over 2}(x_1^2+x_2^2) \qquad
      M^2(Z) = {g^2+g_2^2 \over 2}(x_1^2+x_2^2)
      \ee 
      and
      \bea
      \Psi_1(W^{\pm}) = \Psi_2(W^{\pm}) &=& 
      {3g_2^2 \over 32 \pi^2} f\Bigl(M^2(W^{\pm})\Bigr) \qquad\\[3.5mm]
      \Psi_1(Z) = \Psi_2(Z) &=& 
      {3(g^2+g_2^2) \over 64\pi^2} f\Bigl(M^2(Z)\Bigr) \qquad\nonumber 
      \ena

\item [$\bullet$ Leptons] $(\ell : e,\mu,\tau)$
      \be
      M^2({\ell}) = (Y_{\ell}\,x_1)^2
      \ee
      and
      \be
      \Psi_1(\ell) = -{Y_{\ell}^2 \over 8\pi^2} f\Bigl(M^2(\ell)\Bigr)
      \qquad\qquad \Psi_2(\ell) = 0 
      \ee

\item [$\bullet$ Up-Quarks] $({\cal Q} : u,c,t)$
      \be
      M^2({\cal Q}) = (Y_{{\cal Q}}\,x_2)^2
      \ee
      and
      \be
      \Psi_1({\cal Q}) = 0 \qquad\qquad \Psi_2({\cal Q}) = 
      -{3 Y^2_{{\cal Q}} \over 8\pi^2} f\Bigl(M^2({\cal Q})\Bigr)
      \ee

\item [$\bullet$ Down-Quarks] $(q : d,s,b)$
      \be
      M^2(q) = (Y_{q}\,x_1)^2
      \ee
      and
      \be
      \Psi_1(q) = -{3 Y^2_{q} \over 8\pi^2} f\Bigl(M^2(q)\Bigr) 
      \qquad\qquad \Psi_2(q) = 0
      \ee

\item [$\bullet$ Gluinos] \mbox{}
      \be
      M^2(\tilde{G}) = M^2_3 \qquad\Longrightarrow\qquad \Psi_1(\tilde{G})=
          \Psi_2(\tilde{G})=0  
      \ee
      However, with a ``vacuum energy'' subtraction ($\Omega')$ their 
      contribution is canceled from $V_{1-loop}$.

\item [$\bullet$ Charginos] $(C_1,C_2)$  \\
      Field dependent mass matrix:
      \be
      {\cal M}^T_C {\cal M}_C = \left(
      \begin{array}{cc}
      M_2^2 + g_2^2 x_1^2       &   -g_2(\mu x_1 + M_2 x_2)  \\
      -g_2(\mu x_1 + M_2 x_2)   &   \mu^2 + g_2^2 x_2^2
      \end{array}
      \right)
      \ee
      Its eigenvalues $M^2(C_k)=\lambda_k(x_1,x_2),~k=1,2 $ 
      are roots of the characteristic equation~: 
      \be
      \lambda^2 - b_1\lambda +b_0 =0 
      \ee
      with
      $b_1=\mu^2 + M_2^2 + g_2^2(x_1^2+x_2^2)  \quad\mbox{and}\quad
      b_0 = (\mu M_2 - g_2^2x_1x_2)^2 $. Obviously $b_0,b_1>0$ so
      there are two real and positive roots. Easily :
      \be
      {1 \over x_i}{\partial\lambda_k \over \partial x_i} =
      {1 \over 2}\left[
      {1 \over x_i}{\partial b_1 \over \partial x_i} +
      { {b_1 \over x_i}{\partial b_1 \over \partial x_i}-
        {2 \over x_i}{\partial b_0 \over \partial x_i} \over
      {\rm sign}(\lambda_k - \lambda_{\ell \not= k}) {\cal D}_C}
      \right]  \quad (i,\ell=1,2)
      \ee
      where 
      \be
      {\cal D}_C=\sqrt{b_1^2-4 b_0}=
      \sqrt{(\mu+M_2)^2 + g_2^2 (x_1 - x_2)^2}\;
      \sqrt{(\mu-M_2)^2 + g_2^2 (x_1 +x_2)^2}
      \ee
      Finally
      \be
      \Psi_i(C_k) = -{1 \over 16\pi^2}\,{1 \over x_i}
      {\partial\lambda_k \over \partial x_i}\,f(\lambda_k)
      \ee

\item [$\bullet$ Neutralinos] $(N_1,N_2,N_3,N_4)$  \\
      Field dependent mass matrix:
      \be
      {\cal M}_N = \left(
      \begin{array}{cccc}
      M_1  &  0    &  g x_1/\sqrt{2}     &  -g x_2/\sqrt{2}  \\
      0    &  M_2  &  -g_2 x_1/\sqrt{2}  &  g_2 x_2/\sqrt{2}  \\
      g x_1/\sqrt{2}   &  -g_2 x_1/\sqrt{2}  &  0  &  -\mu  \\
      -g x_2/\sqrt{2}  &  g_2 x_2/\sqrt{2}   &  -\mu  &  0  \\
      \end{array}
      \right)
      \ee
      Mass eigenvalues $M(N_k)=\lambda_k(x_1,x_2),~k=1,2,3,4 $
      are roots of the characteristic equation~:
      \be
      \label{neu:ch}
      \lambda^4 + A_1\lambda^3 + A_2\lambda^2 + A_3\lambda + A_4 = 0
      \ee
      It is straightforward to show that 
      \bea
      A_1 & = & -M_1 - M_2  \\
      A_2 & = & -\mu^2 + M_1 M_2 - (g^2+g_2^2)(x_1^2+x_2^2)/2 \nonumber\\
      A_3 & = & \mu^2(M_1+M_2) - \mu(g^2+g_2^2)x_1x_2 + 
            (M_1g_2^2+M_2g^2)(x_1^2+x_2^2)/2  \nonumber\\
      A_4 & = & -\mu^2 M_1M_2 + \mu (M_1g_2^2+M_2g^2)x_1x_2 \nonumber
      \ena
      Differentiation of Eq.~(\ref{neu:ch}) gives
      \bea
      {1 \over x_i}{\partial\lambda_k \over \partial x_i} &  =  & 
      -{\vartheta_{ik} \over \vartheta'_k} 
      \qquad (i=1,2)\\[3mm]
      \vartheta_{ik} &  =  & 
       {1 \over x_i}{\partial A_1 \over \partial x_i}\lambda_k^3
        +{1 \over x_i}{\partial A_2 \over \partial x_i}\lambda_k^2
        +{1 \over x_i}{\partial A_3 \over \partial x_i}\lambda_k
        +{1 \over x_i}{\partial A_4 \over \partial x_i}    \nonumber \\[3mm]
      \vartheta'_k &  =  & 
      4\lambda_k^3 + 3 A_1\lambda_k^2 + 2 A_2\lambda_k + A_3  \nonumber
      \ena
      Finally
      \be
      \Psi_i(N_k)  =  -{1 \over 32\pi^2}\,{1 \over x_i}
      {\partial\lambda_k^2 \over \partial x_i}\,f(\lambda_k^2)
      \ee

\item [$\bullet$ Sneutrinos] $(\tilde{\nu}_{\ell}) \qquad (\ell : e,\mu,\tau)$ 
      \mbox{}
      \be
      M^2(\tilde{\nu}_{\ell}) = m^2_{\tilde{L}} + {1 \over 4}\,
      (g^2+g_2^2)(x_1^2-x_2^2) 
      \ee
      and
      \be
      \Psi_1(\tilde{\nu}_{\ell}) = -\Psi_2(\tilde{\nu}_{\ell}) = 
      {g^2+g_2^2 \over 64\pi^2}\, f\Bigl(M^2(\tilde{\nu}_{\ell})\Bigr) 
      \ee

\item [$\bullet$ Other scalars] $(\varphi_1, \varphi_2) \quad$ All other
      scalars can be grouped into independent subsets. These are: 
      Sleptons $(\tilde{\ell}_1,\tilde{\ell}_2)$, 
      Up-Squarks $(\tilde{\cal Q}_1,\tilde{\cal Q}_2)$,
      Down-Squarks $(\tilde{q}_1,\tilde{q}_2)$ and 
      Higgses $(H^+,H^-),~(h,H),~(\phi_1,\phi_2)$.  \\
      The field dependent mass matrix, for each one subset, 
      has the generic form
      \be
      {\cal M}^2_{\varphi}=\left(
      \begin{array}{cc}
      P  &  Q   \\
      Q  &  R
      \end{array}
      \right)
      \ee
      where $\varphi$ runs on the previously mentioned pairs and
      the necessary quantities for each scalar subset are given in
      Table~II and III\@.
      In order to find its mass eigenvalues  
      $M^2(\varphi_k)=\lambda_k(x_1,x_2),~k=1,2$ 
      we have to solve the characteristic equation
      \be
      \lambda^2-(P+R)\lambda + PR-Q^2 =0
      \ee
      Differentiation of the above equation gives
      \be
      {1 \over x_i}{\partial\lambda_k \over \partial x_i} =
      {1 \over 2}\left[
      {1 \over x_i}{\partial (P+R) \over \partial x_i} +
      { {(P-R) \over x_i}{\partial (P-R) \over \partial x_i}-
        {4 Q \over x_i}{\partial Q \over \partial x_i} \over
      {\rm sign}(\lambda_k - \lambda_{\ell \not= k}) 
      {\cal D}_{\varphi} }
      \right]  \quad (i,\ell=1,2)
      \ee
      where
      \be
      {\cal D}_{\varphi} = \sqrt{(P-R)^2 + 4 Q^2}
      \ee
      while the radiative corrections become
      \bea
      \Psi_i(\varphi_k) = {1 \over 64\pi^2}\,C(\varphi_k){\cal N}(\varphi_k)
      {1 \over x_i} {\partial\lambda_k \over \partial x_i}\,f(\lambda_k)
      \ena
\end{description}
%-----------------------------------------------------------------------------
%\newpage
\section{Mass eigenstates at specific directions}
%-----------------------------------------------------------------------------
\def\theequation{B.\arabic{equation}}
\setcounter{equation}{0}
%-----------------------------------------------------------------------------
We present below mass eigenvalues and relevant quantities mentioned in 
Sec.~III\@. Note that Yukawa couplings of the
$3^{\rm rd}$ family are denoted, as usual, by 
$(Y_t, Y_b, Y_{\tau}) \equiv (Y_u^3, Y_b^3, Y_e^3)$ and
$M_S$ is the characteristic scale of SUSY breaking,
which is typically an average of the relevant soft parameters.
(Consult also Sec.~II and Appendix~A).
%-----------------------------------------------------------------------------
\subsection{Direction $(\cah_1,\cah_2=\cah_1) \equiv (r\gg,~\theta=\pi/4)$}
%-----------------------------------------------------------------------------
\def\theequation{B.\arabic{equation}}
\setcounter{equation}{0}
%-----------------------------------------------------------------------------
\begin{description}
\item [$\bullet$ Gauge Bosons ($S=1$)] \mbox{}
      \be
      M^2(Z) = {g^2+g_2^2 \over 2}\, r^2 \qquad 
      M^2(W^{\pm}) = {g_2^2\, r^2 \over 2}
      \ee
\item [$\bullet$ Fermions ($S=1/2$)] \mbox{}
      \be
      \begin{array}{lll}
      \dsty M^2(t) = {1 \over 2}\,Y_t^2 r^2 \qquad &  
      \dsty M^2(b)={1 \over 2}\,Y_b^2 r^2  \qquad &  
      \dsty M^2(\tau)={1 \over 2}\,Y_{\tau}^2 r^2     \\[3mm]
      \dsty M^2(C_1) = {g_2^2\, r^2 \over 2} \qquad & 
      \dsty M^2(C_2) = {g_2^2\, r^2 \over 2} &  
      \end{array}
      \ee
      For the neutralino eigenproblem we have assumed a perturbative 
      solution \cite{martin-ram} of the form  
      \[
      \begin{array}{c}
      M(N_i)=\lambda_i \qquad (i=1,2,3,4) \qquad \mbox{with}  \\
      \lambda_1=\xi_1 r \qquad \lambda_2=\xi_2 r \qquad 
      \lambda_3=\xi_3 \qquad \lambda_4=\xi_4 
      \end{array}
      \]
      Since $\lambda_i$ are roots of the characteristic equation
      $\lambda^4+A_1\lambda^3+A_2\lambda^2+A_3\lambda+A_4 =0$ 
      (see Appendix A), we know from elementary algebra that 
      $\lambda_i$ satisfy the following system of equations
      \begin{eqnarray*}
      \lambda_1 + \lambda_2 + \lambda_3 + \lambda_4 
      & = & -A_1 \\
      \lambda_1 \lambda_2 + \lambda_2 \lambda_3 + \lambda_3 \lambda_4 +
      \lambda_4 \lambda_1 + \lambda_1 \lambda_3 + \lambda_2 \lambda_4 
      & = & A_2 \\
      \lambda_1 \lambda_2 \lambda_3 + \lambda_2 \lambda_3 \lambda_4 +
      \lambda_1 \lambda_2 \lambda_4 + \lambda_1 \lambda_3 \lambda_4
            & = & -A_3 \\
      \lambda_1 \lambda_2 \lambda_3 \lambda_4 & = & A_4
      \end{eqnarray*}
      By comparing the highest order terms in $r$ we imediately
      conclude that
      \[     -\xi_1=\xi_2=\sqrt{ {g^2+g_2^2 \over 2} }\qquad   \]
      whereas $\xi_3$, $\xi_4$ are solutions of 
      \[   
      \begin{array}{c}
      -\hat{g}^2 \xi^2 + (\chi - \mu \hat{g}^2) \xi + \mu \chi = 0 , \\
      \hat{g}^2 = g^2+g_2^2 \quad\mbox{and}\quad 
      \chi=M_1 g_2^2+M_2 g^2 
      \end{array}
      \]
      So the relevant eigenvalues become
      \be
      M^2(N_1)=M^2(N_2)={g^2+g_2^2 \over 2}\, r^2 \qquad 
      M^2(N_3)= {\cal O}(M_S^2) \qquad M^2(N_4)= {\cal O}(M_S^2)
      \ee
\item [$\bullet$ Higgses ($S=0$)] Solving the eigenvalue problem to
      highest order in $r$ we get
      \be
      \begin{array}{ll}
      \dsty M^2(H^+)= {g_2^2\, r^2 \over 2} \qquad & \dsty M^2(H^-)=
      {m_1^2+m_2^2 \over 2} - {(m_1^2 - m_2^2)^2 \over 2 g_2^2 r^2}\\[3mm]
      \dsty M^2(H)= {g^2+g_2^2 \over 2}\, r^2 \qquad & \dsty M^2(h)=
      {m_1^2+m_2^2 \over 2} - {(m_1^2 - m_2^2)^2 \over 2 (g^2 + g_2^2) r^2}
      \\[3mm]
      M^2(\phi_1)= {\cal O}(M_S^2) \qquad & M^2(\phi_2)= {\cal O}(M_S^2)
      \end{array}
      \ee
\item [$\bullet$ Super scalars ($S=0$)] \mbox{}
      \be
      \begin{array}{lll}
      \dsty M^2(\tilde{\nu}_3) = {\cal O}(M_S^2) \qquad &
      \dsty M^2(\tilde{\tau}_1)={1 \over 2}\,Y_{\tau}^2 r^2 \qquad & 
      \dsty M^2(\tilde{\tau}_2)= {1 \over 2}\,Y_{\tau}^2 r^2   \\[3mm]
      \dsty M^2(\tilde{t}_1)= {1 \over 2}\,Y_t^2 r^2 \qquad &
      \dsty M^2(\tilde{t}_2)= {1 \over 2}\,Y_t^2 r^2 &  \\[3mm]
      \dsty M^2(\tilde{b}_1)= {1 \over 2}\,Y_b^2 r^2 \qquad &
      \dsty M^2(\tilde{b}_2)= {1 \over 2}\,Y_b^2 r^2 &
      \end{array}
      \ee
\end{description}
%-----------------------------------------------------------------------------
\subsection{ Direction $(0,\cah_2) \equiv (r\gg,~\theta=\pi/2)$ }
%-----------------------------------------------------------------------------
\begin{description}
\item [$\bullet$ Gauge Bosons ($S=1$)] \mbox{} 
      \be
      M^2(Z) = {g^2+g_2^2 \over 2}\, r^2 \qquad 
      M^2(W^{\pm}) = {g_2^2\, r^2 \over 2}
      \ee
\item [$\bullet$ Fermions ($S=1/2$)] \mbox{} 
      \be
      \begin{array}{lll}
      M^2(t) = Y_t^2 r^2 \qquad & M^2(b)=0 \qquad & 
      M^2(\tau)=0   \\[3mm]
      M^2(C_1) = g_2^2\, r^2 \qquad & \dsty M^2(C_2)=
                 {\mu^2 M^2_2 \over g_2^2\, r^2 }&  
      \end{array}
      \ee
      For the neutralino sector as before, we assume a perturbative 
      solution now with a slightly different  form
      \[
      \lambda_1=\xi_1 r \qquad \lambda_2=\xi_2 r \qquad 
      \lambda_3=\xi'_3 \qquad  \lambda_4=\xi'_4/r^2 
      \]
      Solving to highest order we obtain 
      \[ 
      -\xi_1=\xi_2=\sqrt{ {g^2+g_2^2 \over 2} }\qquad 
      \xi'_3={M_1 g_2^2+M_2 g^2 \over g^2+g_2^2}\qquad
      \xi'_4={2 \mu^2 M_1 M_2 \over M_1 g_2^2 + M_2 g^2}
      \]
      and the relevant eigenvalues become
      \be
      M^2(N_1)=M^2(N_2)={g^2+g_2^2 \over 2}\, r^2 \qquad 
      M^2(N_3)={\cal O}(M_S^2)  \qquad
      M^2(N_4)={{\xi'_4}^2 \over r^4}
      \ee
\item [$\bullet$ Higgses ($S=0$)] \mbox{}
      \be
      \begin{array}{ll}
      \dsty M^2(H^+)={g^2+g_2^2 \over 4}\, r^2 \qquad & 
      \dsty M^2(H^-)= {-g^2+g_2^2 \over 4}\, r^2 \\[3mm]
      \dsty M^2(H)={3(g^2+g_2^2) \over 4}\, r^2 \qquad & 
      \dsty M^2(h)= -{g^2+g_2^2 \over 4}\, r^2  \\[3mm]
      \dsty M^2(\phi_1)= -{g^2+g_2^2 \over 2}\, r^2 \qquad & 
      \dsty M^2(\phi_2)= {g^2+g_2^2 \over 2}\, r^2
      \end{array}
      \ee
\item [$\bullet$ Super scalars ($S=0$)] \mbox{}
      \be
      \begin{array}{ll}
      \dsty M^2(\tilde{\nu}_3) = -{g^2+g_2^2 \over 4} \qquad & \\[3mm]
      \dsty M^2(\tilde{\tau}_1)={g^2 r^2 \over 2} \qquad & 
      \dsty M^2(\tilde{\tau}_2)= {-g^2+g_2^2 \over 4}\, r^2  \\[3mm]
      \dsty M^2(\tilde{t}_1)= Y_t^2 r^2 - {g^2 \over 3}\, r^2 \qquad &
      \dsty M^2(\tilde{t}_2)= Y_t^2 r^2 + {g^2-3 g_2^2 \over 12}\, r^2  
      \\[3mm]
      \dsty M^2(\tilde{b}_1)={g^2 r^2 \over 6} \qquad &
      \dsty M^2(\tilde{b}_2)={g^2+3 g_2^2 \over 12}\, r^2   
      \end{array}
      \ee
\end{description}
Finally, the expressions (3.2) for the one-loop effective potential 
along $(0,\cah_2)$ direction require the following quantities
\be
\begin{array}{llll}
\dsty U_1={g^2+g_2^2 \over 2} \qquad & 
\dsty U_2={g_2^2 \over 2} \qquad & U_3=g_2^2  \qquad & 
\dsty U_4={g^2+g_2^2 \over 4} \\[3mm]
\dsty U_5={g^2 \over 2} \qquad   & 
\dsty U_6={-g^2+g_2^2 \over 2} \qquad &
\dsty U_7={g^2 \over 6} \qquad & \\[3mm]
\dsty U_8={g^2+3g_2^2 \over 12} &
\dsty U_9={3(g^2+g_2^2) \over 4} \qquad & 
\dsty U_{10}=Y_t^2 - {g^2 \over 3} \qquad & 
\dsty U_{11}=Y_t^2 +{g^2 -3g_2^2 \over 12} 
\end{array}
\ee
and
\be
   \vec{d} \equiv (1,~6,~-4,~5,~2,~4,~6,~6,~1,~6,~6)    
\ee
%-----------------------------------------------------------------------------
%\newpage
\section{Realization for $\beta$-function thresholds}
%-----------------------------------------------------------------------------
\def\theequation{C.\arabic{equation}}
\setcounter{equation}{0}
%-----------------------------------------------------------------------------
The generic form of a step function appearing in MSSM RGEs with one-loop
thresholds \cite{sak1} is
\be
     \theta_i = \theta\Bigl(\tilde{Q}^2 - T_i(\cah_1,\cah_2)\Bigr)
     \qquad(i=1,2,\ldots,22)
\ee
where $T_i$ defines the decoupling point. Following the notation of 
Sec.~II and Appendix~A, we shall present here an 
implementation of the various $T_i$ we have used in our investigation.

\noindent
$\bullet$ Field configurations near EW breaking minimum ($ \alt 300$ GeV)
\be
\begin{array}{lllll}
T_1 = M_1^2 \qquad&   T_2 = M_2^2 \qquad&   T_3 = M_3^2 \qquad&   
T_4 = m_{H_1}^2 \qquad&   T_5 = m_{H_2}^2 \\[3mm]
T_6 = \mu^2 \qquad&   T_7 = \mu^2 \qquad&   T_8 = m^2_{\tilde{Q}_3}  \qquad&   
T_9 = m^2_{\tilde{Q}_2}  \qquad&   T_{10} = m^2_{\tilde{Q}_1}  \\[3mm]
T_{11} = m^2_{\tilde{U}^c_3} \qquad&   T_{12} = m^2_{\tilde{U}^c_2} \qquad&   
T_{13} = m^2_{\tilde{U}^c_1} \qquad&   T_{14} = m^2_{\tilde{D}^c_3} \qquad&   
T_{15} = m^2_{\tilde{D}^c_2}  \\[3mm]
T_{16} = m^2_{\tilde{D}^c_1} \qquad&   T_{17} = m^2_{\tilde{L}_3} \qquad&   
T_{18} = m^2_{\tilde{L}_2} \qquad&   T_{19} = m^2_{\tilde{L}_1} \qquad&   
T_{20} = m^2_{\tilde{E}^c_3}  \\[3mm]
T_{21} = m^2_{\tilde{E}^c_2} \qquad&   T_{22} = m^2_{\tilde{E}^c_1}
\end{array}
\ee
$\bullet$ Otherwise\footnote
%-----------------------------------------------------------------------------
{We use $M^2({\cal P})$ to denote the field dependent mass eigenstates of 
the particle ${\cal P}$.}
%-----------------------------------------------------------------------------
\be
\begin{array}{llll}
T_1 = |M^2(N_3)| \qquad  &  T_2 = |M^2(C_2)| \qquad  &  
T_3 = |M_3^2| \qquad   &   T_4 = |M^2(\phi_1)| \qquad  \\[3mm]
%-----------------------------------------------------------------------------
T_5 = |M^2(H^+)|  &  T_6 = |M^2(N_4)| \qquad  &   T_7 = |M^2(C_1)| \qquad  &
T_8 = |M^2(\tilde{t}_1)|  \qquad  \\[3mm]
%-----------------------------------------------------------------------------
T_9 = |M^2(\tilde{c}_1)|  \qquad  &   T_{10} = |M^2(\tilde{u}_1)|  \qquad  &
T_{11} = |M^2(\tilde{t}_2)| \qquad  &   T_{12} = |M^2(\tilde{c}_2)| \\[3mm]
%-----------------------------------------------------------------------------
T_{13} = |M^2(\tilde{u}_2)| \qquad  &   T_{14} = |M^2(\tilde{b}_1)| \qquad  &
T_{15} = |M^2(\tilde{s}_1)|  \qquad  &  T_{16} = |M^2(\tilde{d}_1)| \\[3mm]
%-----------------------------------------------------------------------------
T_{17} = |M^2(\tilde{\tau}_1)| \qquad  &   
T_{18} = |M^2(\tilde{\mu}_1)| \qquad  &   
T_{19} = |M^2(\tilde{e}_1)| \qquad  &   T_{20} = |M^2(\tilde{\tau}_2)| \\[3mm]
%-----------------------------------------------------------------------------
T_{21} = |M^2(\tilde{\mu}_2)| \qquad  &   T_{22} = |M^2(\tilde{e}_2)|
\end{array}
\ee
%-----------------------------------------------------------------------------
%\newpage
\section{Numerical issues}
%-----------------------------------------------------------------------------
\def\theequation{D.\arabic{equation}}
\setcounter{equation}{0}
%-----------------------------------------------------------------------------
\noindent
$\bullet$ {\bf Rounding error:}
Squeezing infinitely many real numbers into a finite number of bits
(binary digits) requires an approximate representation. Given any 
fixed number of bits, (e.g., Double Precision has 64 bits) most 
calculations with real numbers will produce 
quantities that can not be exactly represented using that many bits.
Therefore, the result of a floating-point calculation must often be 
rounded in order to fit back into its finite representation. This 
rounding error is the characteristic feature of floating-point 
computation. 

Estimation of this kind of error is in general an extremely
complicated process. However, for the trivial operations ($+,~-,~*,~/$) 
one can give bounds for the rounding errors involved \cite{stoer,wilk},
From our perspective, the computation of $V^{(1)}$,
in case of huge field dependent (FD) mass eigenvalues, might suffer from 
large rounding error. This error could arise from subtractions of numbers 
that are nearly equal or additions and subtractions of numbers that differ 
greatly in magnitude.  Clearly, if it is greater in absolute value
than $V^{(0)}$, then our resulted $V_{1-loop}$ would be untrustworthy.
One can show \cite{wilk} that a crude upper bound for the relevant rounding 
error due to summation is
\be
\begin{array}{c}
\displaystyle
    E_S(V^{(1)}) \alt \delta \sum_k |V_k^{(1)}|,\qquad\qquad
      \delta = {1 \over 2} 10^{1-t}   \\
  \mbox{For Double Precision accuracy $t=16$}
\end{array}
\ee
Fortunately enough, one can reduce rounding error in these cases, by 
grouping the operands according to relative size, so that as much as 
possible operations will be performed between numbers with similar 
magnitudes.  The recommended method is sorting the partial sums 
($V^{(1)}_k$) with respect to their absolute value before summing 
them.\footnote
%-----------------------------------------------------------------------------
{This procedure gives the smallest upper bound in $E_S$, but it does
not necessarily give the smallest error in practice.}
%-----------------------------------------------------------------------------
Afterwards, summation is performed by using the Kahan formula \cite{gold}.
Consequently, we have to restrict ourselves only to those field 
configurations for which
\be
\label{err}
   |V^{(0)}| > E_S(V^{(1)})
\ee
is satisfied. Indeed, from our numerical study it seems that for all 
$|\cah_1|,~|\cah_2| < 10^{11}$ GeV Eq.~(\ref{err}) holds. So one should 
not worry for this kind of error.

\noindent\\
$\bullet$ {\bf Catastrophic canselation error:}
The evaluation of any expression, containing an addition of quantities 
with opposite signs, could result in a relative error so large that all 
the digits are meaningless. When subtracting nearby quantities the most 
significant digits in the operands match and cancel each other.

Especially, a catastrophic canselation occurs when the operands are 
approximately equal and subject to rounding errors. However, a formula 
that exhibits 
catastrophic canselation can sometimes be arranged to eliminate the 
problem.  We have faced this problem in computing  some FD 
mass eigenstates of various MSSM sectors. These, for 
most MSSM scalar particles, are determined by solving the eigenvalue 
problem of a generic $2 \times 2$ mass matrix 
\be
   {\cal M}_{\varphi} = \left(
         \begin{array}{cc}
           P & Q \\
           Q & P 
         \end{array}
                    \right) 
\ee
The solution of its characteristic equation 
\be
\label{eigen}
\begin{array}{c}
         x^2+b x+c =0 \\
         b=P+Q,\qquad c=PQ-R^2
\end{array}
\ee
involves the computation of two ``dangerous'' quantities 
\be
\Delta= b^2-4c \qquad\mbox{and}\qquad Z_{\pm}=-b \pm \sqrt{b^2-4c}
\ee
When $b^2 \approx 4c$ catastrophic canselation\footnote
%-----------------------------------------------------------------------------
{The quantities $b^2$ and $4c$ are subject to rounding errors since
they are the results of floating point multiplications.}
%-----------------------------------------------------------------------------
can cause many of the accurate digits to disappear leaving behind 
mainly digits contaminated by rounding error.\footnote
%-----------------------------------------------------------------------------
{Indeed we found that in some cases round off error completely ruins
the results by turning $\Delta$ into an imaginary quantity.}
%-----------------------------------------------------------------------------
Luckily enough, this shortcoming can be fixed by rearranging the terms
of $\Delta$. Namely, we can use the following expression
\be
   \Delta = (P+Q)^2 -4 (PQ-R^2)=(P-Q)^2+4 R^2
\ee
everywhere in our computations. On the other hand, if $c$
is small then one of $Z_{\pm}$ will involve the subtraction
of $b$ from a very nearly equal quantity ($\Delta$) and the associated
root comes out with large inaccuracy. However, a definitely more accurate
way to compute the roots of (\ref{eigen}) is found in \cite{num-rec}
\be
\begin{array}{c}
  q=-{1 \over 2}\left[ b+ {\rm sign}(b)\sqrt{b^2-4c}\; \right] \\[3mm]
\displaystyle
      x_1= q \qquad\mbox{and}\qquad x_2={c \over q}
\end{array}
\ee
The expression $x^2-y^2$ is another formula that exhibits catastrophic
cancelation (when $x \approx y$). To compute $x^2-y^2$, as accurate as 
possible, we apply \cite{stoer}
\be
      x^2-y^2 = \cases{
      (x+y)(x-y),       &  if $~~~~{1 \over \sqrt{3} } < 
            \left| {x \over y} \right| < \sqrt{3} $ \cr
       x x-y y          &  otherwise           \cr}
\ee

\noindent\\
$\bullet$ {\bf Diagonalization:}
The optimum strategy for finding eigenvalues is first to reduce the 
matrix to a simple form
and afterwards begin an iterative procedure. The most efficient 
program for finding all eigenvalues of a symmetric matrix, is a 
combination of the Householder tridiagonalization and the QL algorithm 
with implicit shifts (for details see \cite{num-rec,reinch}).
Note that when dealing with a matrix, whose elements vary over many
orders of magnitude, it is important that the smallest elements are in 
the top left-hand corner.  This is because the reduction is performed 
starting from the bottom right-hand corner and a mixture of small and 
large elements there can lead to considerable rounding errors.
One possible way to overcome this difficulty is to perform a trivial 
reflection through secondary diagonal when matrix elements are not 
properly aligned. Of course this is an orthogonal 
transformation so the eigenvalues are not affected.

\noindent\\
$\bullet$ {\bf Minimization aspects:}
Minimization of the effective potential $V_{1-loop}$ was performed
by running the ``merlin'' software package on
a SGI Origin 2000 supercomputer. Merlin is an integrated environment
designed to solve optimization problems of the following form :
\begin{quote}
Find a local minimum of the function 
\[
    f(\vec{x}),\quad \vec{x}\in{\bf R}^N,\quad 
      \vec{x}=[x_1,x_2,\ldots,x_N]^T
\]
under the conditions 
\[
      x_i \in [a_i,b_i]\quad \mbox{for}~i=1,2,\ldots,N
\]
\end{quote}
It contains implementations of powerful minimization algorithms.
In particular, there are two direct methods (ROLL, SIMPLEX) that use
no derivative information and three algorithms from the conjugate 
gradient family (Fletcher-Reevs, Polar-Ribiere and the generalized 
Polar-Ribiere). Also from the Quasi-Newton methods the DFP and several 
versions of BFGS method are coded.

To improve minimization's effectiveness,
we have made a linear change of independent variables (scaling) so 
that the values of the new variables at the minimum are of order unity
\cite{min-guru}.
For convenience, we have also imposed an upper bound on the 
domain of $V_{1-loop}$ coding in our routines the following potential 
function\footnote
%-----------------------------------------------------------------------------
{Our choise is $S_{\infty}\simeq 10^{11}$ GeV and 
$V_{Big}=10^{300}~({\rm GeV})^4$.}
%-----------------------------------------------------------------------------
\be
\label{sgi}
      V_{SGI} = \cases{
      V_{1-loop},      &  if $~~~~|\cah_1| ,|\cah_2| < S_{\infty}$\cr
       V_{Big}         &  otherwise           \cr}
\ee
We have tried many minimization algorithms to our potential function.
Of course, due to lack of analytic derivatives and the extremely 
``flat'' behaviour of (\ref{sgi}), near H-diag direction, direct 
methods have an advantage over the gradient ones.
Indeed, by means of the SIMPLEX algorithm we were able to scan the 
H-diag direction in its entierty for probable local minima.
%-----------------------------------------------------------------------------
%                             REFERENCES
%-----------------------------------------------------------------------------

%-----------------------------------------------------------------------------
%                               TABLES
%-----------------------------------------------------------------------------
\newpage
\begin{table}
\caption{Experimental bounds on supersymmetric particles (in units of GeV).}
\begin{tabular}{l||cccc}
Neutralinos & $M_{N_1} > 10.9$  &  $M_{N_2} > 45.3$  &  $M_{N_3} > 75.8$  &  
$M_{N_4} > 127$  \\ 
\hline
Charginos &   &  $M_{C_1} > 65.7$  &  $M_{C_2} > 99$  &   \\ 
\hline
Sneutrino-Higgses~~& $M_{\tilde{\nu}} > 43.1$  & 
$M_h >62.5$  &  $M_{H} >77.5$  &  $M_{H^{\pm}} > 54.5$     \\ 
\hline
Squarks-Sleptons & $M_{\tilde{q}} > 224$ & $M_{\tilde{e}} > 58$  & 
$M_{\tilde{\mu}} > 55.6$  & $M_{\tilde{\tau}} > 45$
\end{tabular}
\end{table}
%-----------------------------------------------------------------------------
\begin{table}
\caption{Degrees of freedom for all MSSM mass eigenstates in a neutral
         Higgs background.}
\begin{tabular}{lc||cc|cc|cc}
\multicolumn{2}{c||}{Eigenstate} & Spin (S)& & Color (C)& & 
Helicity (${\cal N}$) & \\ 
\hline\hline
Gauge Bosons & $(W^{\pm})$ & 1 & & 1 & & 2 &  \\
%\cline{2-8}
            & $(Z)$ & 1 & & 1 & & 1 &  \\
\hline
Leptons & $(\ell : e,\mu,\tau)$ & 1/2 & & 1 & & 2 &  \\
\hline
Up-Quarks & $({\cal Q} : u,c,t)$ & 1/2 & & 3 & & 2 &  \\
\hline
Down-Quarks & $(q : d,s,b)$ & 1/2  & & 3 & & 2 &  \\
\hline
Gluinos &  &  1/2  & &  8 & & 1 & \\
\hline
Charginos & $(C_1,C_2)$ & 1/2 & & 1 & & 2 &   \\
\hline
Neutralinos & $(N_1,N_2,N_3,N_4)$ & 1/2 & & 1 & & 1 &  \\
\hline
Sneutrinos & $(\tilde{\nu}_{\ell})$ & 0 & & 1 & & 2 & \\ 
\hline
Sleptons & $(\tilde{\ell_1}, \tilde{\ell_2})$ & 0 & & 1 & & 2 & \\ 
\hline
Up-Squarks & $(\tilde{\cal Q}_1,\tilde{\cal Q}_2)$ & 0 & & 3 & & 2 & \\ 
\hline
Down-Squarks & $(\tilde{q}_1,\tilde{q}_2)$ & 0 & & 3 & & 2  & \\ 
\hline
Higgses & $(H^+,H^-)$ & 0 & & 1 & & 2 & \\ 
%\cline{2-8}
       & $(h,H)$     & 0 & & 1 & & 1 & \\ 
%\cline{2-8}
       & $(\phi_1, \phi_2)$ & 0 & & 1 & & 1 & 
\end{tabular}
\end{table}
%-----------------------------------------------------------------------------
\begin{table}
\caption{Mass matrix entries for some MSSM eigenstates
         in a neutral Higgs background.}
\begin{tabular}{c||c|c|c}
Eigenstate   &   P   &   Q   &   R  \\
\hline\hline
Sleptons & &  & \\
$(\tilde{\ell_1}, \tilde{\ell_2})$ &
$ m^2_{\tilde{E}^c} + Y_{\ell}^2 x_1^2 + {g^2 \over 2}(x_2^2-x_1^2)$ & 
$ Y_{\ell}(A x_1+\mu x_2)$  &  
$ m^2_{\tilde{L}} + Y^2_{\ell} x_1^2 + {g^2-g_2^2 \over 4}(x_1^2-x_2^2)$ \\
\hline
Up-Squarks & &  & \\
$(\tilde{\cal Q}_1,\tilde{\cal Q}_2)$ &
$m^2_{\tilde{U}^c} + Y^2_{\cal Q} x_2^2 - {g^2 \over 3}(x_2^2-x_1^2)$ &
$-Y_{\cal Q}(\mu x_1+ A x_2)$ &
$m^2_{\tilde{Q}} + Y^2_{\cal Q} x_2^2 + {g^2-3 g_2^2 \over 12}(x_2^2-x_1^2)$ \\
\hline
Down-Squarks & & & \\
$(\tilde{q}_1,\tilde{q}_2)$ &
$m^2_{\tilde{D}^c} + Y^2_q x_1^2 + {g^2 \over 6}(x_2^2-x_1^2)$ &
$Y_q(A x_1+ \mu x_2)$ &
$m^2_{\tilde{Q}} + Y^2_q x_1^2 +{g^2+3 g_2^2 \over 12}(x_2^2-x_1^2)$ \\
\hline
Higgses & & & \\
$(H^+,H^-)$ &
$m_1^2 +{g^2+g_2^2 \over 4}(x_1^2-x_2^2) + {g_2^2 \over 2}x_2^2$ &
$-m_3^2 + {g_2^2 \over 2}x_1 x_2$ &
$m_2^2 + {g^2+g_2^2 \over 4}(x_2^2-x_1^2) + {g_2^2 \over 2}x_1^2$ \\
$(h,H)$     &
$m_1^2 +{g^2+g_2^2 \over 4}(3 x_1^2-x_2^2)$  &
$m_3^2 - {g^2 +g_2^2 \over 2}x_1 x_2$  &
$m_2^2 + {g^2+g_2^2 \over 4}(3 x_2^2-x_1^2)$  \\
 $(\phi_1,\phi_2)$ &
$m_1^2 +{g^2+g_2^2 \over 4}(x_1^2-x_2^2)$ &
$-m_3^2$ &
$m_2^2 + {g^2+g_2^2 \over 4}(x_2^2-x_1^2)$  
\end{tabular}
\end{table}
%-----------------------------------------------------------------------------
%                            FIGURE CAPTIONS
%-----------------------------------------------------------------------------
\newpage
\thispagestyle{empty}
\begin{figure}
\centering 
\vspace*{5cm}
\psfig{figure=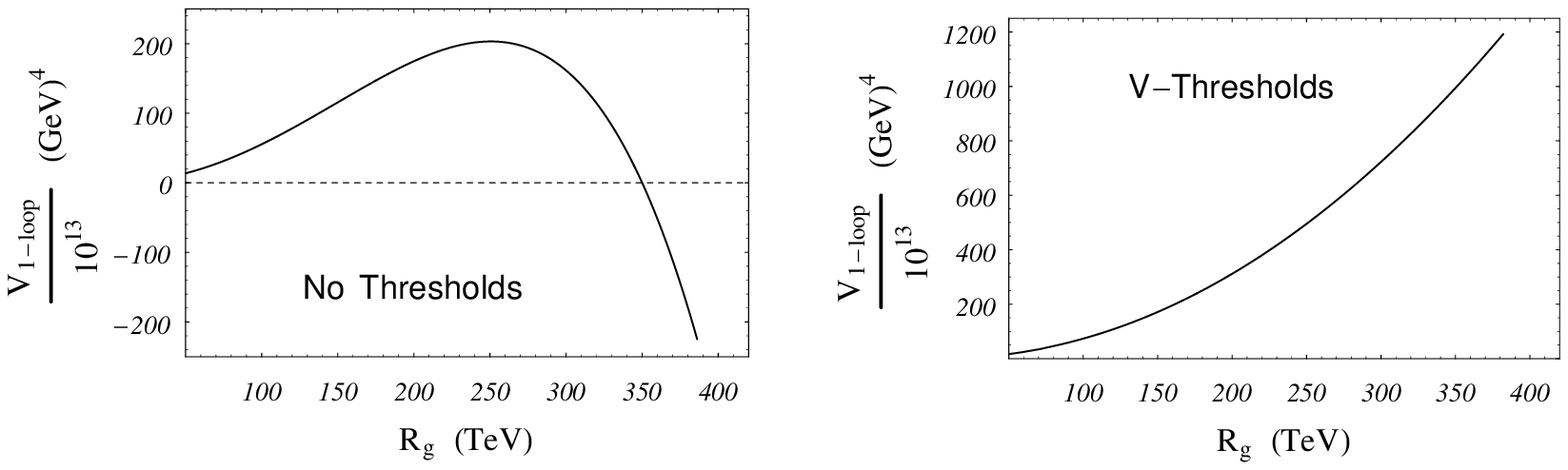,height=5cm,width=15cm} 
\vspace{0.5cm}
(a)~~~~~~~~~~~~~~~~~~~~~~~~~~~~~~~~~~~~~~~~~~~~~~~~~~~(b)
\vspace{5cm}
\caption[]{ The 1-loop potential as a function of the ``generalized radious''
$R_g(\phi_1) = \sqrt{\phi_1^2+\phi_2^2}$ is shown in the 
$A_0=-400$ GeV, $M_0=60$ GeV, $M_{1/2}=200$ GeV and $\tan\beta=2$
case. $(\phi_1,\phi_2(\phi_1))$ is the lowest point of H-Diag 
direction at an arbitrary field value $\phi_1=\cah_1(Q_{high})$.  }
\end{figure}
%-----------------------------------------------------------------------------
\newpage
\def\thefigure{\arabic{figure}a}
%-----------------------------------------------------------------------------
\thispagestyle{empty}
\begin{figure}
\centering
\vspace*{5cm}
\psfig{figure=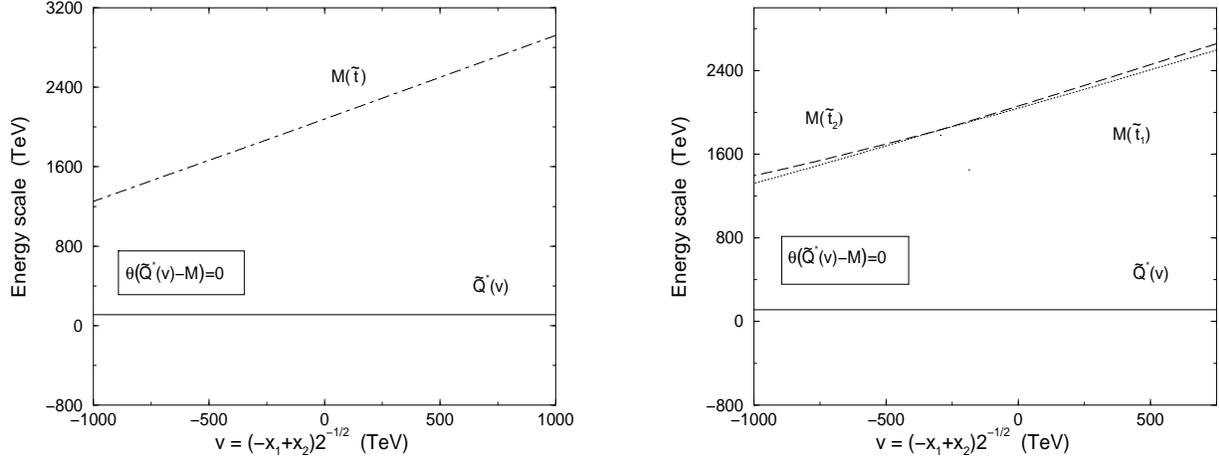,height=6cm,width=16cm}
\vspace{5cm}
\caption[]{ Decoupling process of the top-stops field dependent mass 
eigenvalues as we are moving along the line
$x_1+x_2=\phi_1+\phi_2(\phi_1)$, where $x_i = \cah_i$.
$(\phi_1,\phi_2(\phi_1))$ represents the lowest point of H-Diag direction 
when $\phi_1=\cah_1(Q_{high}) \simeq 2 \times 10^3$ TeV 
(initial conditions at $M_X$ same as in Fig.~1). }
\end{figure}
%-----------------------------------------------------------------------------
\def\thefigure{\arabic{figure}b}
\setcounter{figure}{1}
%-----------------------------------------------------------------------------
\thispagestyle{empty}
\begin{figure}
\centering
\vspace*{5cm}
\psfig{figure=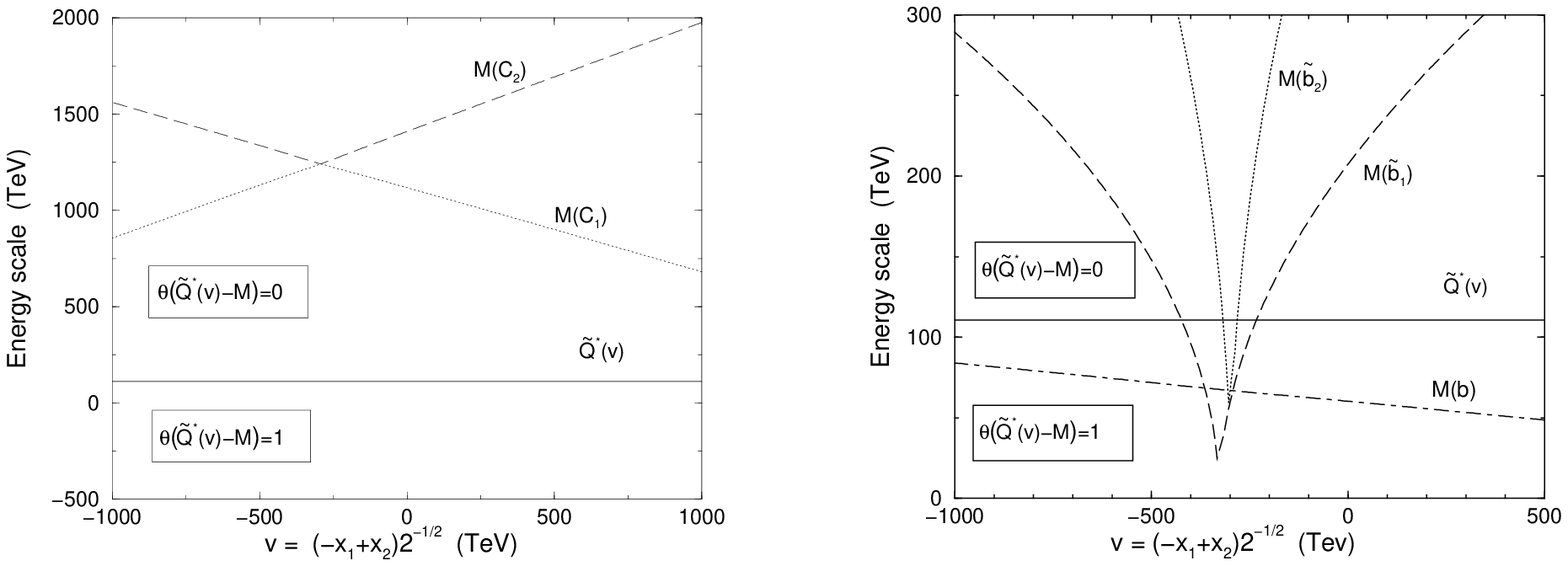,height=6cm,width=16cm}
\vspace{5cm}
\caption[]{ Same as Fig.~2a but now for charginos and bottom-sbottoms.}
\end{figure}
%-----------------------------------------------------------------------------
\def\thefigure{\arabic{figure}a}
%-----------------------------------------------------------------------------
\thispagestyle{empty}
\begin{figure}
\centering
\vspace*{2.5cm}
\psfig{figure=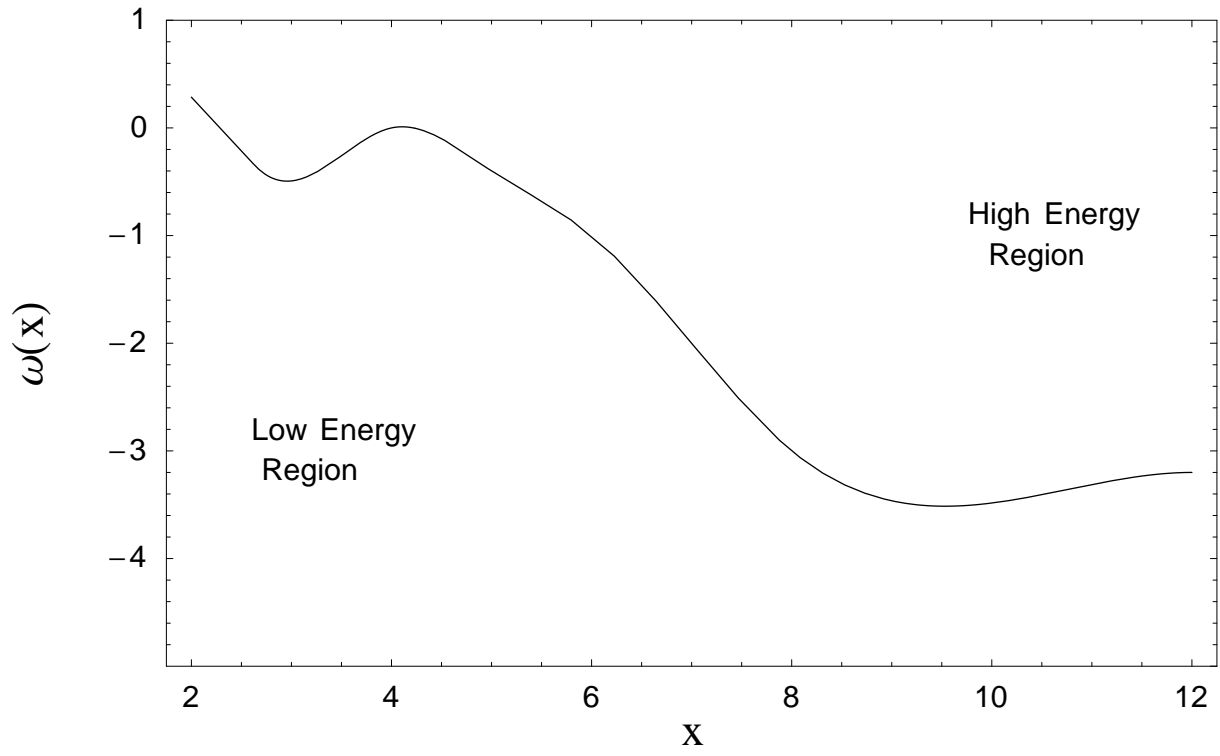,width=16cm}
\vspace{5cm}
\caption[]{Plot of the $\omega(x)$ used in the definition of the
critical scale $\tilde{Q}^*$. }
\end{figure}
%-----------------------------------------------------------------------------
\def\thefigure{\arabic{figure}b}
\setcounter{figure}{2}
%-----------------------------------------------------------------------------
\thispagestyle{empty}
\begin{figure}
\centering
\vspace*{5cm}
\psfig{figure=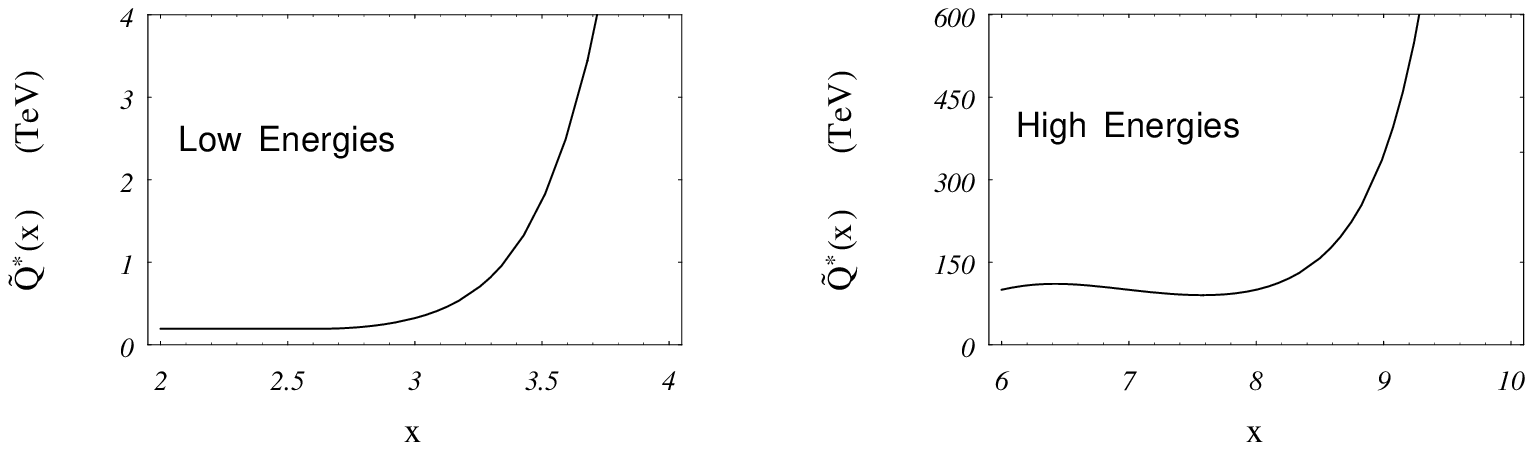,height=5cm,width=16cm}
\vspace{5cm}
\caption[]{The renormalization scale used for RG improvement of 
the effective potential as a function of 
$x={1 \over 2}\log{\cah_1^2+\cah_2^2 \over L^2}.$ }
\end{figure}
\thispagestyle{empty}
%-----------------------------------------------------------------------------
%                                THE END
%-----------------------------------------------------------------------------
\end{document}